\documentclass[11pt]{article}

\usepackage[utf8]{inputenc}
\usepackage[T1]{fontenc}
\usepackage[a4paper,margin=1in]{geometry}
\usepackage{ragged2e}
\usepackage{microtype}
\microtypesetup{expansion=false}
\usepackage{color}
\usepackage{graphicx}
\usepackage{amsmath}
\usepackage{amssymb}
\usepackage{booktabs}
\usepackage{orcidlink}
\usepackage{cite}
\usepackage{subcaption}
\usepackage{verbatim}
\usepackage{tabularx}
\usepackage{placeins}
\hypersetup{hidelinks}

\justifying 

\begin{document}
	
	\title{From the Early Slope to Curvature: Future Prospects and challenges for Astrophysical Parameter Estimation Using the Core-Collapse Supernova High-Frequency Feature}
	
	\author{%
		Alejandro Casallas-Lagos\orcidlink{0000-0001-5985-8819}\thanks{Corresponding author: \href{mailto:a.casallas-lagos@uw.edu}{a.casallas-lagos@uw.edu}}\\
		\small Faculty of Physics, University of Warsaw, Warsaw, Poland
		\and
		Marek Szczepa\'nczyk\orcidlink{0000-0002-6167-6149}\\
		\small Faculty of Physics, University of Warsaw, Warsaw, Poland
		\and
		Michele Zanolin\orcidlink{0000-0002-4044-4306}\\
		\small Department of Physics and Astronomy, ERAU, Arizona, U.S.
		\and
		Anthony Mezzacappa\orcidlink{0000-0001-9816-9741}\\
		\small Department of Physics and Astronomy, University of Tennessee, Knoxville, U.S.
	}
	\date{}

	\maketitle
	
	\begin{abstract}
		Gravitational-wave (GW) signals from core-collapse supernovae (CCSNe) contain both stochastic and deterministic components. Among these, the High-Frequency Feature (HFF), associated with Proto-Neutron Star (PNS) oscillations, has emerged as an observable for probing dense-matter physics. Previous studies of the HFF have primarily focused on its early-time slope as a diagnostic of PNS contraction and its dependence on the nuclear equation of state (EOS). In this work, we investigate recovering higher-order features of the HFF evolution, such as its curvature, from core-collapse supernova numerical simulations. By analyzing 2D and 3D waveforms injected into real interferometric noise, we apply the coherent WaveBurst algorithm in its cWB-XP implementation to reconstruct the HFF of recovered events. Comparing reconstructions at different signal strengths allows us to examine the extent to which the HFF remains accessible beyond its early linear evolution
	\end{abstract}

	
	\section{Introduction}
	Numerical-simulation and data-analysis communities have made substantial progress toward characterizing the still-undetected gravitational-wave (GW) signals from core-collapse supernovae (CCSNe). These efforts have established that CCSN GW signals are predominantly stochastic, reflecting the turbulent and nonlinear dynamics of the explosion \cite{Andresen_2017,Couch_2013,Dimmelmeier_2008,Janka_2012,Ott_2013,Radice_2019}. Nevertheless, these signals may also contain deterministic aspects associated with identifiable physical processes in the source \cite{Bizouard_2021,Couch_2013,O_Connor_2011,Ott_2013,Powell_2020,Powell_2022,MeMaLa23,Murphy_2025,Sotani_2021,Radice_2019}.
	
	Among these aspects, a prominent signature present in many modern CCSN simulations is associated with oscillations of the proto-neutron star (PNS), commonly identified with fundamental $f$- and gravity $g$-mode activity \cite{Astone_2018,Cerd_Dur_n_2013,Morozova_2018,M_ller_2013}. In this work, we refer to this structure as the high-frequency feature (HFF); a detailed discussion of this terminology and its physical motivation is provided in \cite{pv6p-dtr2}. In time--frequency representations, the HFF generally appears as a rising ridge that often begins at frequencies of approximately $150$--$250~\mathrm{Hz}$ and may extend toward $\sim1$--$2~\mathrm{kHz}$. During an early portion of its evolution, the ridge can be approximated by a linear frequency evolution.
	
	Physically, the HFF is associated with oscillation modes of the PNS excited by accretion flows and hydrodynamic instabilities, particularly Ledoux convection \cite{Mezz_D_GW_2023, Murphy_2025}. The initial slope of its frequency evolution is sensitive to the contraction of the PNS and therefore to the evolution of its mass, radius, equation of state (EOS),  and neutrino cooling \cite{M_ller_2019,Powell_2022,Morozova_2018,Janka_2012}. The slope should not, however, be interpreted as a direct measurement of the contraction rate alone, because accretion, thermal structure, and relativistic effects also contribute to the observed frequency evolution. Furthermore, the HFF may not correspond to a single $g$-mode throughout its duration, but may instead involve alternating $f$- and $g$-modes \cite{Morozova_2018,Westernacher_2019,Westernacher_2020,Torres_Forn__2019,Murphy_2025}, or multiple $g$-mode branches associated with different regions of the PNS. This possible ambiguity in mode identity is central to the physical interpretation of the feature.
	
	Recent data-analysis efforts have focused primarily on quantifying the early HFF slope under realistic interferometric-noise conditions \cite{Casallas_1,Murphy_2024}. Initial approaches employed chi-squared minimization of low-order polynomial fits to rising frequency tracks reconstructed using the coherent WaveBurst (cWB) burst-detection algorithm \cite{PhysRevD.107.083017}. Complementary methodologies combine normal-mode decomposition \cite{Torres_Forn__2019, Cerd_Dur_n_2013, Bizouard_2021}, polynomial interpolation \cite{Morozova_2018, Astone_2018}, and injections into synthetic Gaussian noise \cite{Bruel_2023}  to investigate the time evolution of the remnant's mass and radius. Collectively, these studies have set the importance of the HFF connecting CCSN GW phenomenology with the evolving structure of the PNS.
	
	Extending the HFF analysis beyond the linear regime may provide complementary sensitivity to the evolving contraction, accretion history, and oscillation modes of the PNS. Section~\ref{sec:hff_curvature} introduces the mathematical framework adopted for this discussion and the limitations that must be addressed before curvature can be established as a reliable inference observable.
	
	The cWB algorithm has been developed to search coherently for a broad range of unmodelled GW transients. In this work, we consider the cWB-XP implementation and its application to the post-processing of CCSNe events reconstructed in realistic interferometric noise \cite{Klimenko_2022}.  A technical description of the reconstruction framework and the theoretical developments still required for beyond-linear HFF measurements in LIGO--Virgo--KAGRA (LVK) data is presented in Section~\ref{sec:cwb}. The present article should therefore be understood as a perspective and roadmap rather than as a completed quantitative curvature-inference study.

	\section{The Core-Collapse Supernova Gravitational-Wave High-Frequency Feature: From Early Slope to Curvature}
	\label{sec:hff_curvature}
	During an early post-bounce interval, approximately $150$--$250~\mathrm{ms}$ for the simulations considered in previous slope studies \cite{Casallas_1,Murphy_2024}, the HFF frequency can be approximated locally as increasing linearly with time. This early evolution is commonly characterized through the initial slope of the feature, primarily using the part of the reconstructed track below approximately $1~\mathrm{kHz}$.
	The frequency evolution around a chosen reference time $t_{\mathrm{s}}$ may be expanded as
		\begin{equation}
			f(t)
			=
			f(t_{\mathrm{s}})
			+
			\dot{f}(t_{\mathrm{s}})(t-t_{\mathrm{s}})
			+
			\frac{1}{2}\ddot{f}(t_{\mathrm{s}})
			(t-t_{\mathrm{s}})^2
			+
			\mathcal{O}\left[(t-t_{\mathrm{s}})^3\right].
			\label{eq:hff_local_expansion}
	\end{equation}
	Here, $t_{\mathrm{s}}$ is a reference time within the interval being characterized, $\dot{f}(t_{\mathrm{s}})$ is the local HFF slope, and $\ddot{f}(t_{\mathrm{s}})$ describes the local rate of change of that slope.
	In this perspective, we use the term \emph{HFF curvature} to refer to the beyond-linear evolution represented locally by the second-order term in equation~\eqref{eq:hff_local_expansion}. Such an equation~\eqref{eq:hff_local_expansion} provides a mathematical description rather than a complete measurement prescription. An estimator of the HFF curvature requires a specified fitting model, frequency range, time interval, and event-selection criterion among others. Establishing and fully characterizing such an estimation process is beyond the scope of the present paper and constitutes an objective for future studies. 
		The physical motivation for extending the characterization beyond the initial slope follows from the nonlinear evolution of the PNS. For illustrative purposes, if we assume that the HFF is given by $f$-mode oscillations and if we are informed by the functional forms used in attempts to fit the HFF (e.g., in pursuit of so-called ``universal relations''), the fundamental frequency may be written as
		\begin{equation}
			f_{\mathrm{HFF}}(t) 
			=
			\sum_{n=0}^{m}
			C_{n}
			\left(
			\sqrt{
				\frac{G M_{\mathrm{PNS}}(t)}
				{R_{\mathrm{PNS}}^{3}(t)}
			}
			\right)^{n}
			\label{eq:hff_frequency_scaling}
	\end{equation}
	where $G$ is the gravitational constant, $M_{\mathrm{PNS}}(t)$ and $R_{\mathrm{PNS}}(t)$ are the time-dependent PNS mass and radius, respectively, $C_{n}$ are constants, and the sum is taken to order $m$. (A logarithmic dependence is also sometimes included in the functional form assumed in equation (\ref{eq:hff_frequency_scaling}), but for our purposes here, the functional form we have assumed in equation (\ref{eq:hff_frequency_scaling}) will suffice.) Taking the logarithmic time derivative of equation~\eqref{eq:hff_frequency_scaling} gives
		\begin{equation}
			\frac{\dot{f}_{\mathrm{HFF}}}{f_{\mathrm{HFF}}}
			=
			\left[
			\sum_{n=1}^{m}
			nC_{n}
			\left(
			\sqrt{
				\frac{G M_{\mathrm{PNS}}(t)}
				{R_{\mathrm{PNS}}^{3}(t)}
			}
			\right)^{n-1}
			\frac{d}{dt}\left(
			\sqrt{
				\frac{G M_{\mathrm{PNS}}(t)}
				{R_{\mathrm{PNS}}^{3}(t)}
			}
			\right)
			\right]
			/
			\sum_{n=0}^{m}
			C_{n}
			\left(
			\sqrt{
				\frac{G M_{\mathrm{PNS}}(t)}
				{R_{\mathrm{PNS}}^{3}(t)}
			}
			\right)^{n},
			\label{eq:hff_frequency_derivative}
	\end{equation}

		where
		\begin{equation}
			\frac{d}{dt}\left(
			\sqrt{
				\frac{G M_{\mathrm{PNS}}(t)}
				{R_{\mathrm{PNS}}^{3}(t)}
			}
			\right)
			=\frac{1}{2}
			\left(
			\frac{G M_{\mathrm{PNS}}(t)}
			{R_{\mathrm{PNS}}^{3}(t)}
			\right)^{-1/2}
			\frac{G M_{\mathrm{PNS}}(t)}
			{R_{\mathrm{PNS}}^{3}(t)}
			\left(
			\frac{\dot{M}}{M}-3\frac{\dot{R}}{R}
			\right).
		\end{equation}
		The HFF relative slope is therefore sensitive to both the relative PNS contraction (i.e., $\dot{R}/R$) and the relative mass accretion (i.e., $\dot{M}/M$).
		Taking a second time derivative provides a connection between the curvature of the HFF and the evolving PNS mass and radius. In particular, differentiating equation (\ref{eq:hff_frequency_derivative})  gives
		\begin{eqnarray}
			\frac{\ddot{f}_{\mathrm{HFF}}}{f_{\mathrm{HFF}}}
			& = &
			\left\{
			\sum_{n=2}^{m}
			n(n-1)C_{n}
			\left(
			\sqrt{
				\frac{G M_{\mathrm{PNS}}(t)}
				{R_{\mathrm{PNS}}^{3}(t)}
			}
			\right)^{n-2}
			\left[\frac{d}{dt}\left(
			\sqrt{
				\frac{G M_{\mathrm{PNS}}(t)}
				{R_{\mathrm{PNS}}^{3}(t)}
			}
			\right)
			\right]^{2}
			\right.
			\nonumber \\
			& &
			\left.
			+
			\sum_{n=1}^{m}
			nC_{n}
			\left(
			\sqrt{
				\frac{G M_{\mathrm{PNS}}(t)}
				{R_{\mathrm{PNS}}^{3}(t)}
			}
			\right)^{n-1}
			\frac{d^{2}}{dt^{2}}
			\left(
			\sqrt{
				\frac{G M_{\mathrm{PNS}}(t)}
				{R_{\mathrm{PNS}}^{3}(t)}
			}
			\right)
			\right\}
			/
			\sum_{n=0}^{m}
			C_{n}
			\left(
			\sqrt{
				\frac{G M_{\mathrm{PNS}}(t)}
				{R_{\mathrm{PNS}}^{3}(t)}
			}
			\right)^{n},
			\label{eq:hff_second_frequency_derivative}
		\end{eqnarray}

		where
		\begin{equation}
			\frac{d^{2}}{dt^{2}}
			\left(
			\sqrt{
				\frac{G M_{\mathrm{PNS}}(t)}
				{R_{\mathrm{PNS}}^{3}(t)}
			}
			\right)
			=
			\frac{GM}{R^{3}}
			\left[
			\frac{\ddot{M}}{M}
			-6\frac{\dot{M}}{M}\frac{\dot{R}}{R}
			-3\frac{\ddot{R}}{R}
			+12\left(\frac{\dot{R}}{R}\right)^{2}
			\right].
			\label{eq:fdoubledotoverf}
		\end{equation}
		This expression shows that the departure of the HFF from the early linear evolution 
		depends on the PNS mass, radius, PNS mass accretion, PNS contraction, and the acceleration or deceleration in the latter two.  
		Equation (\ref{eq:fdoubledotoverf}) is presented here to motivate going beyond the early linear evolution of the HFF.
	
	We remark that the classification of different PNS oscillation modes is currently under investigation by various groups, using different analyses, nomenclatures, simulations, and modeling, also different classifications are adopted across the community. For a discussion regarding these classifications, we encourage the reader to consult \cite{Murphy_2025}. In this work, we do not adopt a specific PNS classification framework from those available in the literature. Instead, we explicitly mention the context and models implemented when necessary, as a detailed discussion on the validity of such classifications is beyond the scope of this work. However, we note that the inclusion of an HFF curvature observable is a natural extension of early-slope studies. This second-order evolution may provide complementary information about PNS and PNS oscillation-mode evolution that is currently inaccessible by examining only the early-slope evolution.
	
	Different physical ingredients should be taken into account when characterizing the HFF curvature, such as the presence of magnetic fields and rotation, even if the presence or absence of rotation can be independently assessed from the core-bounce component in rapidly rotating CCSNe \cite{1jvh-7fdb} among others. For the latter, centrifugal support can reduce the effective gravity and slow PNS contraction, while Coriolis effects modify the oscillation-mode spectrum. The magnitude of these effects depends on the angular-velocity profile and may be parameterized using quantities such as the ratio of rotational kinetic energy to gravitational binding energy, $T_{\mathrm{rot}}/\vert{}W\vert{}$. A centrifugally supported PNS may therefore produce an HFF shifted toward lower frequencies and with a reduced early slope \cite{Pajkos_2019}. This behavior can arise from both a reduction in the restoring force governing the oscillations and a slower evolution of the PNS structure.
	
\begin{figure}
		\centering
		\includegraphics[width=0.30\textwidth]{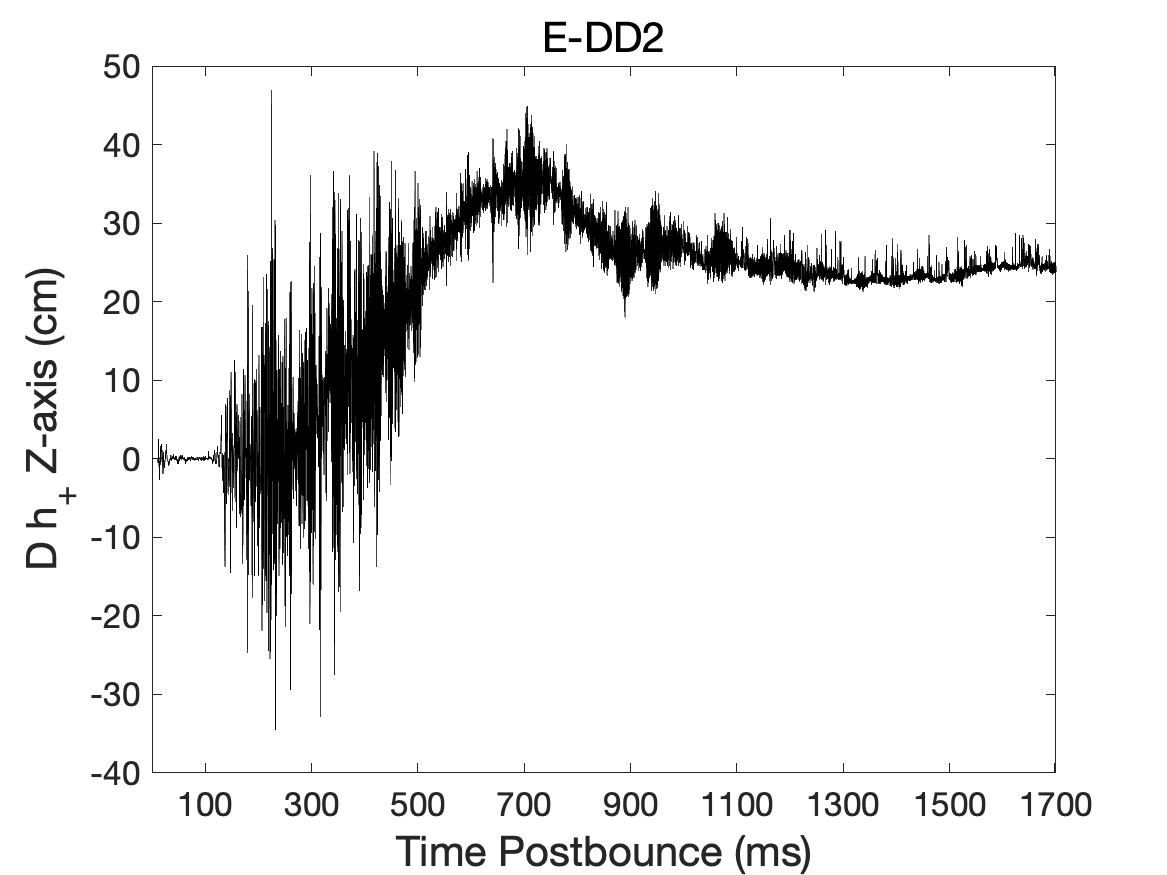}
		\includegraphics[width=0.35\textwidth]{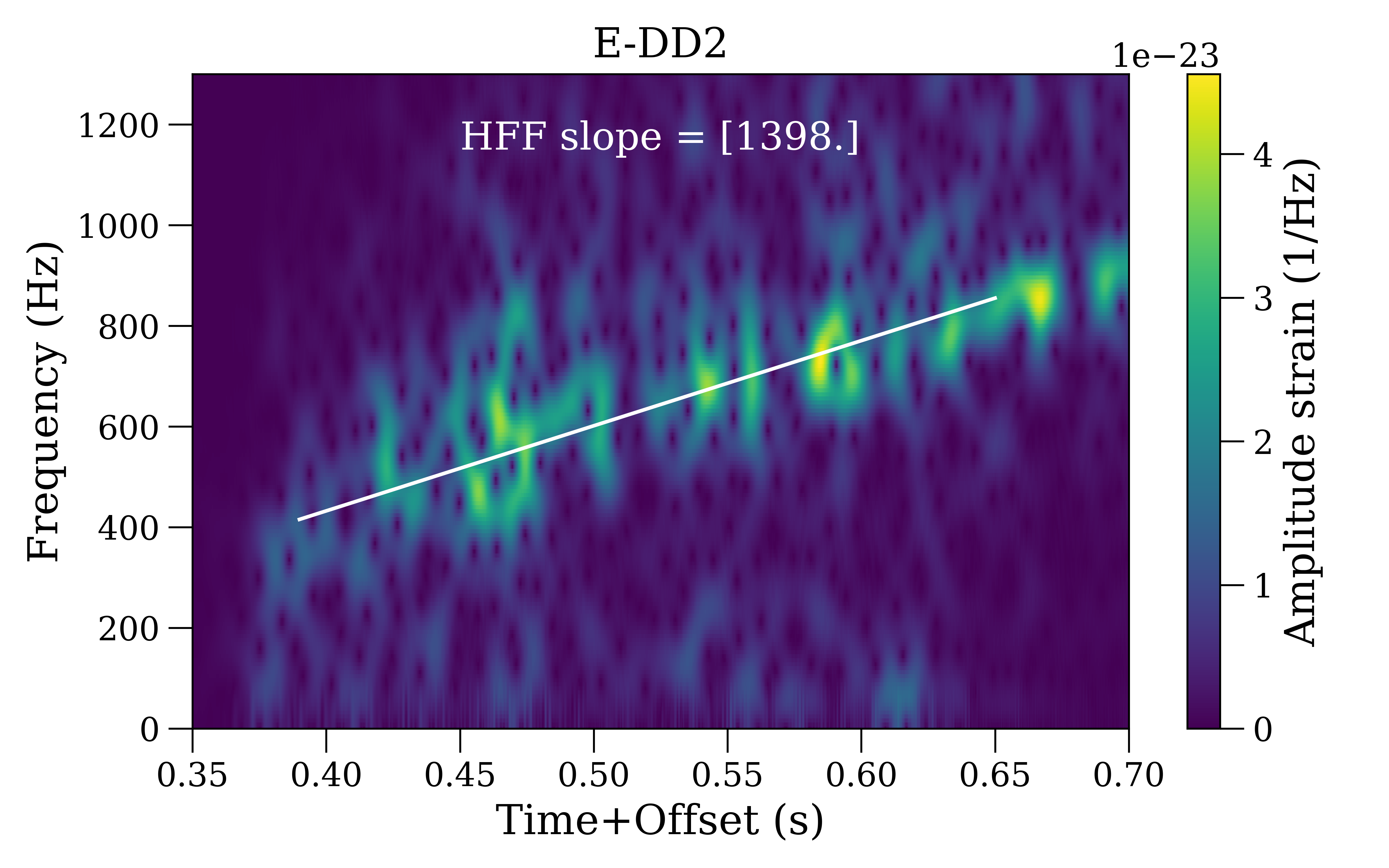}
		\includegraphics[width=0.30\textwidth]{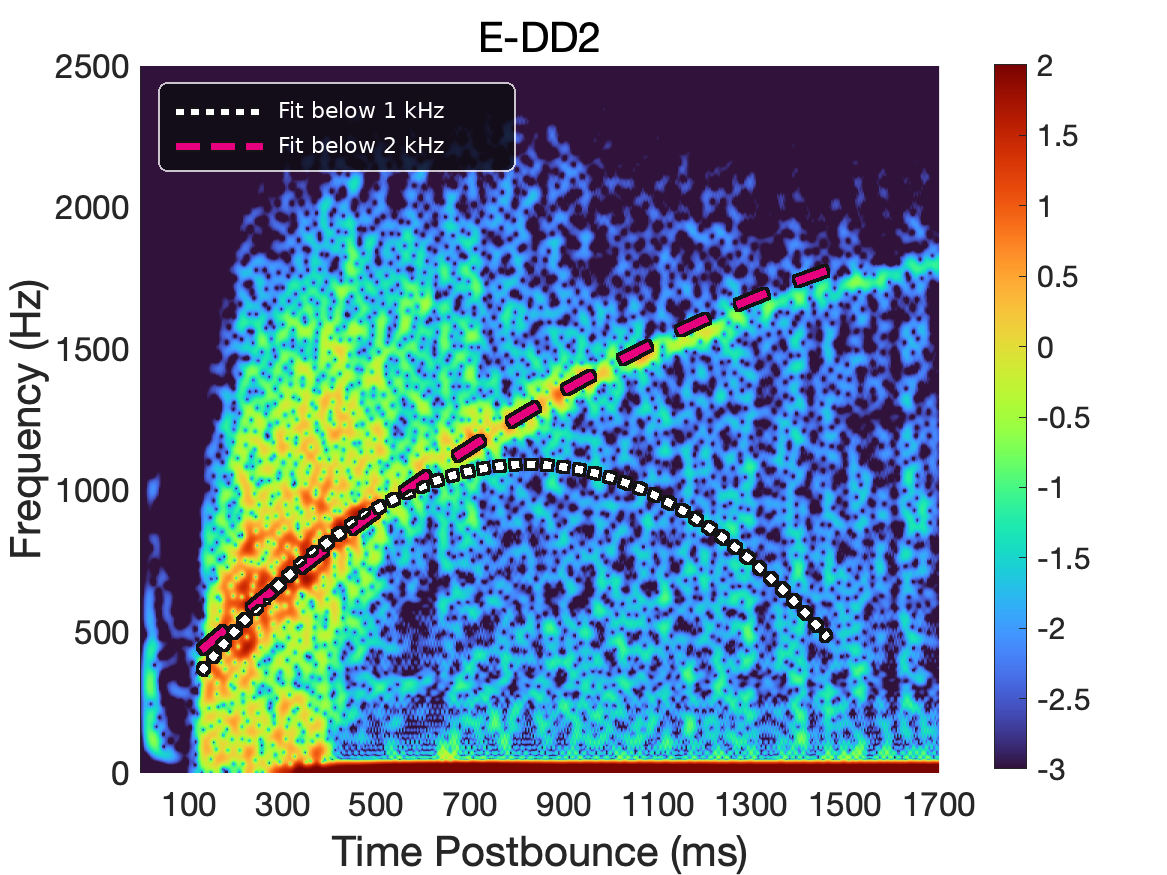}
		\includegraphics[width=0.30\textwidth]{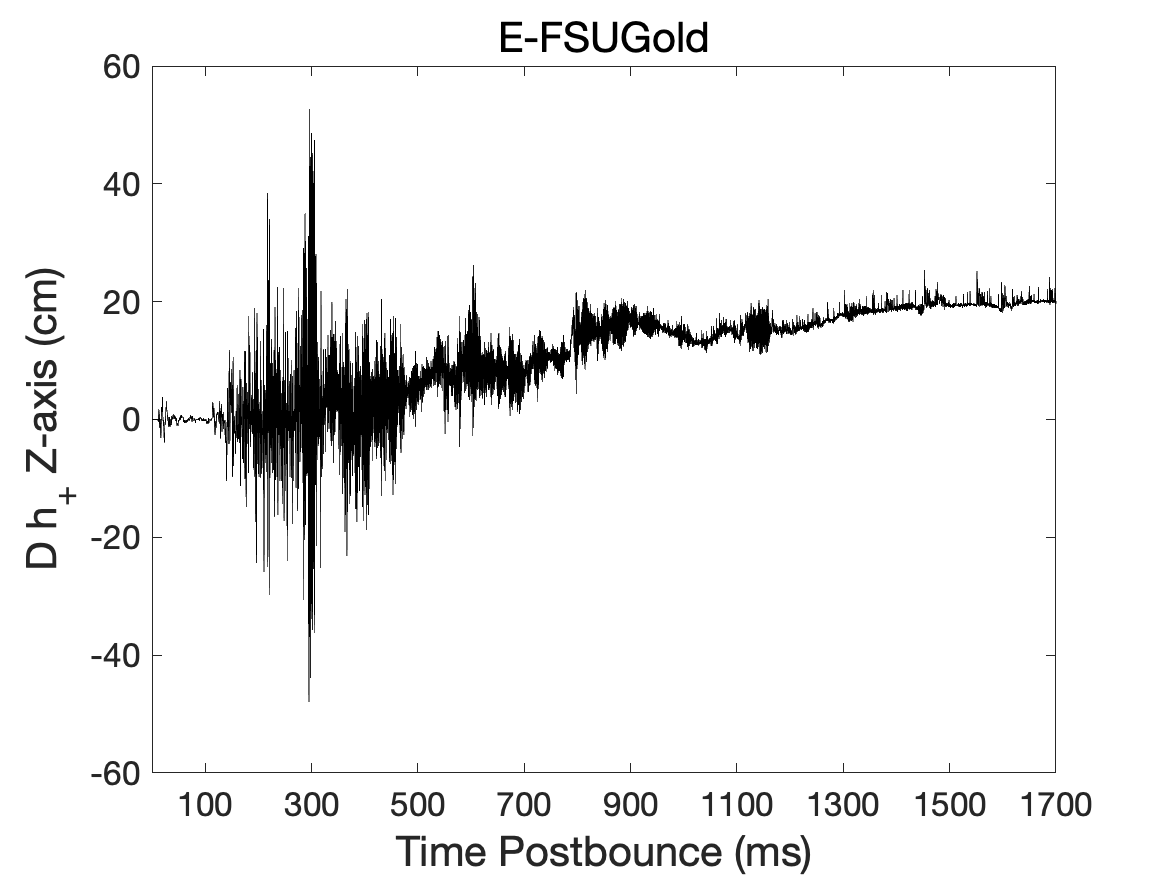}
		\includegraphics[width=0.35\textwidth]{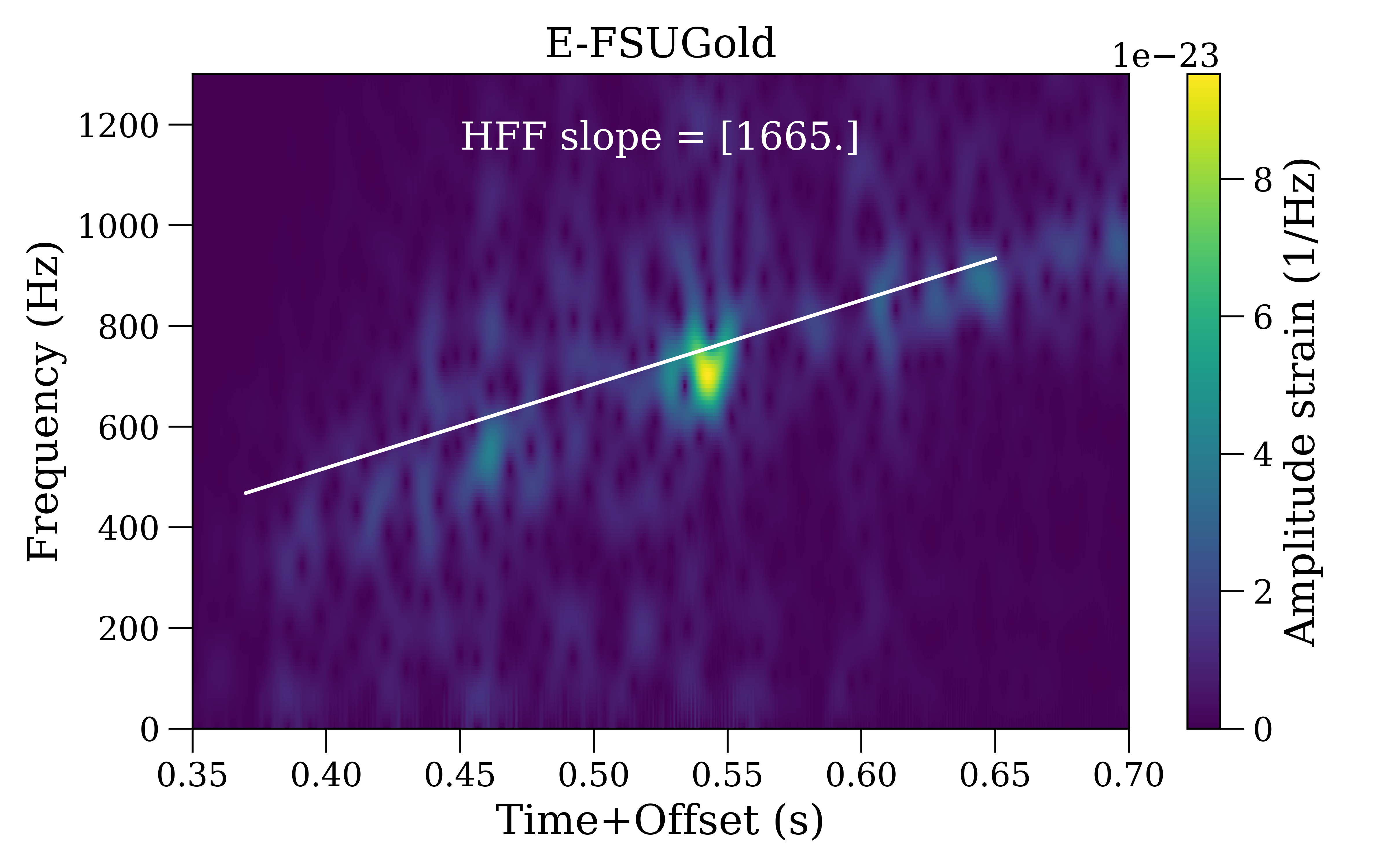}
		\includegraphics[width=0.30\textwidth]{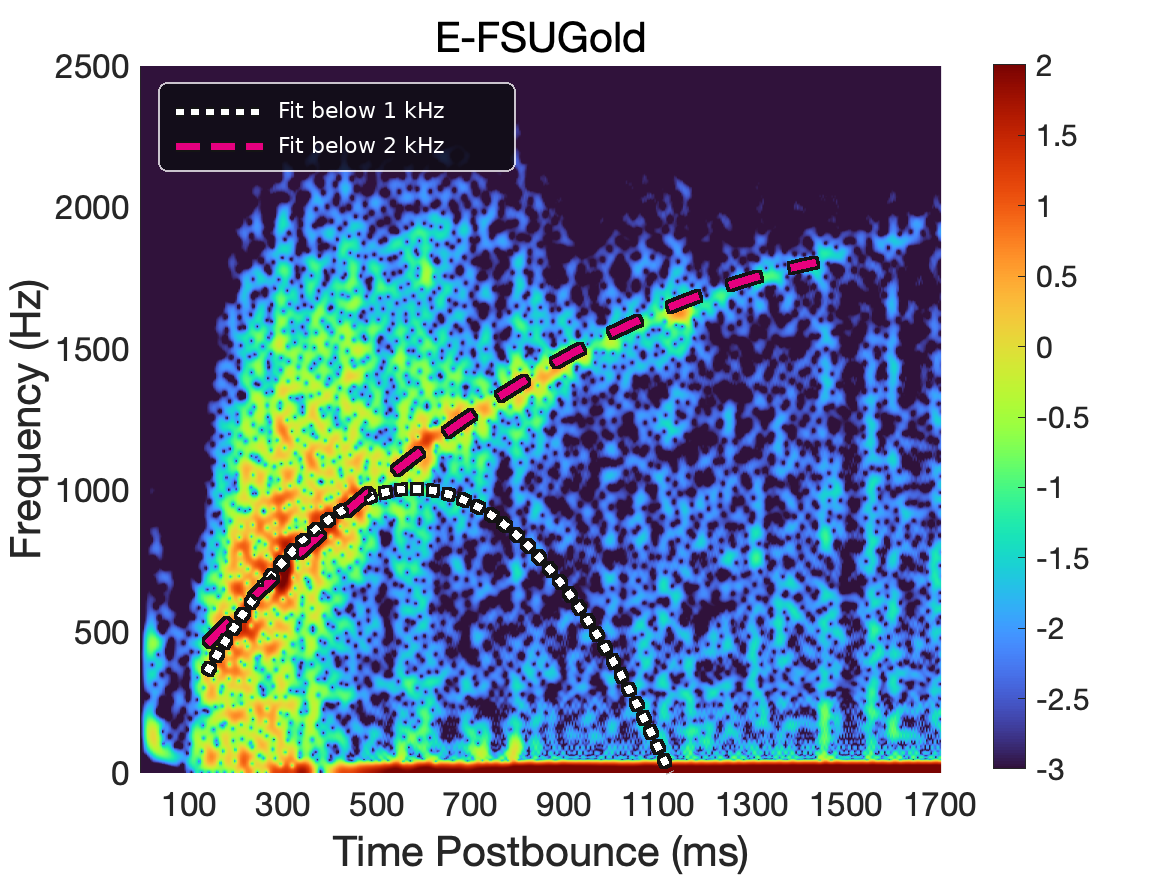}
		\includegraphics[width=0.30\textwidth]{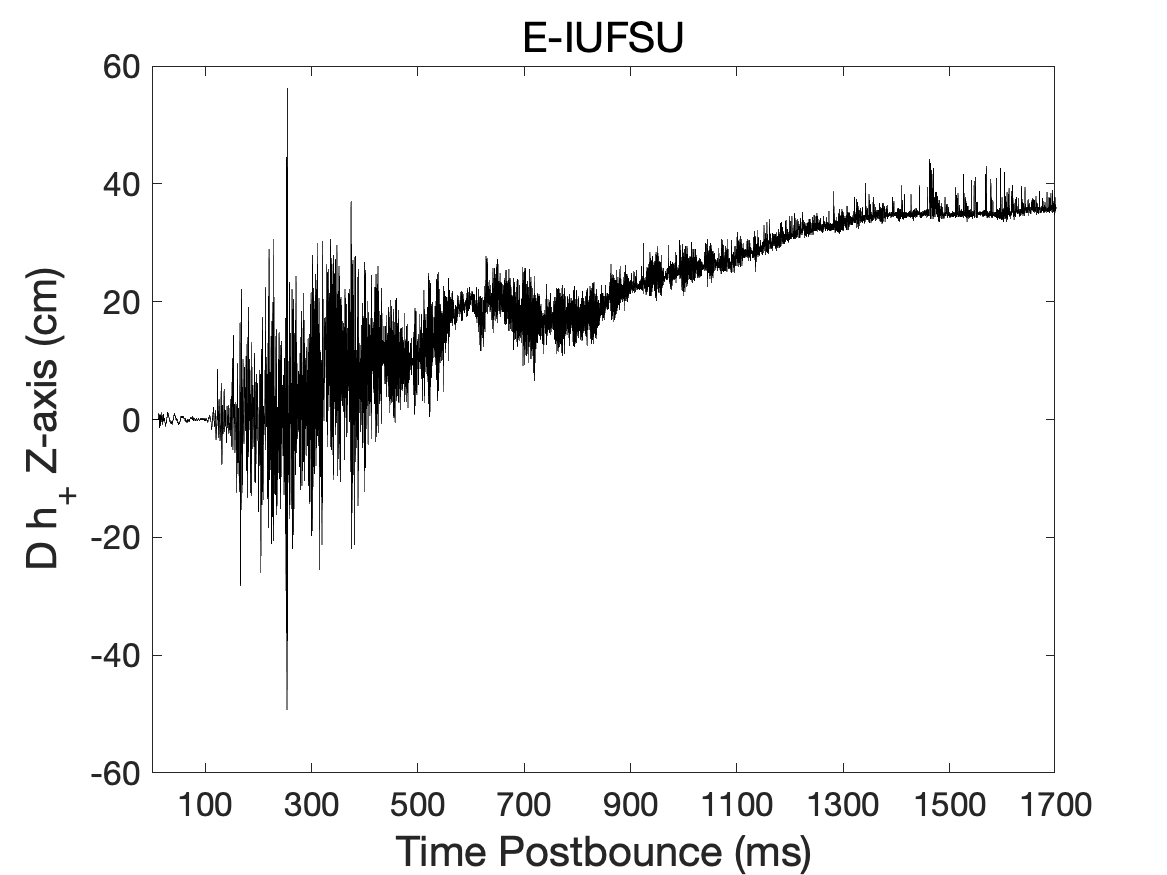}
		\includegraphics[width=0.35\textwidth]{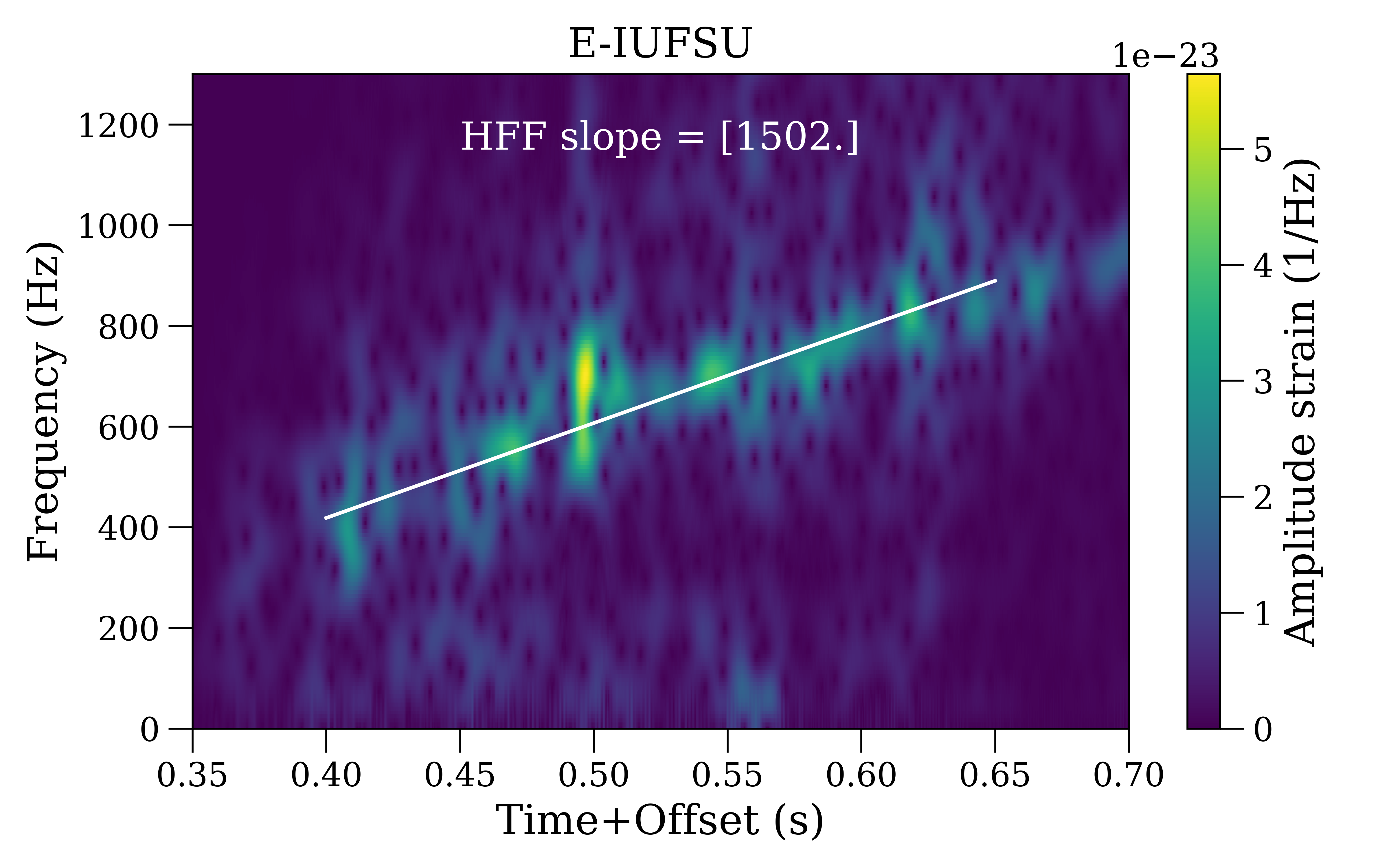}
		\includegraphics[width=0.30\textwidth]{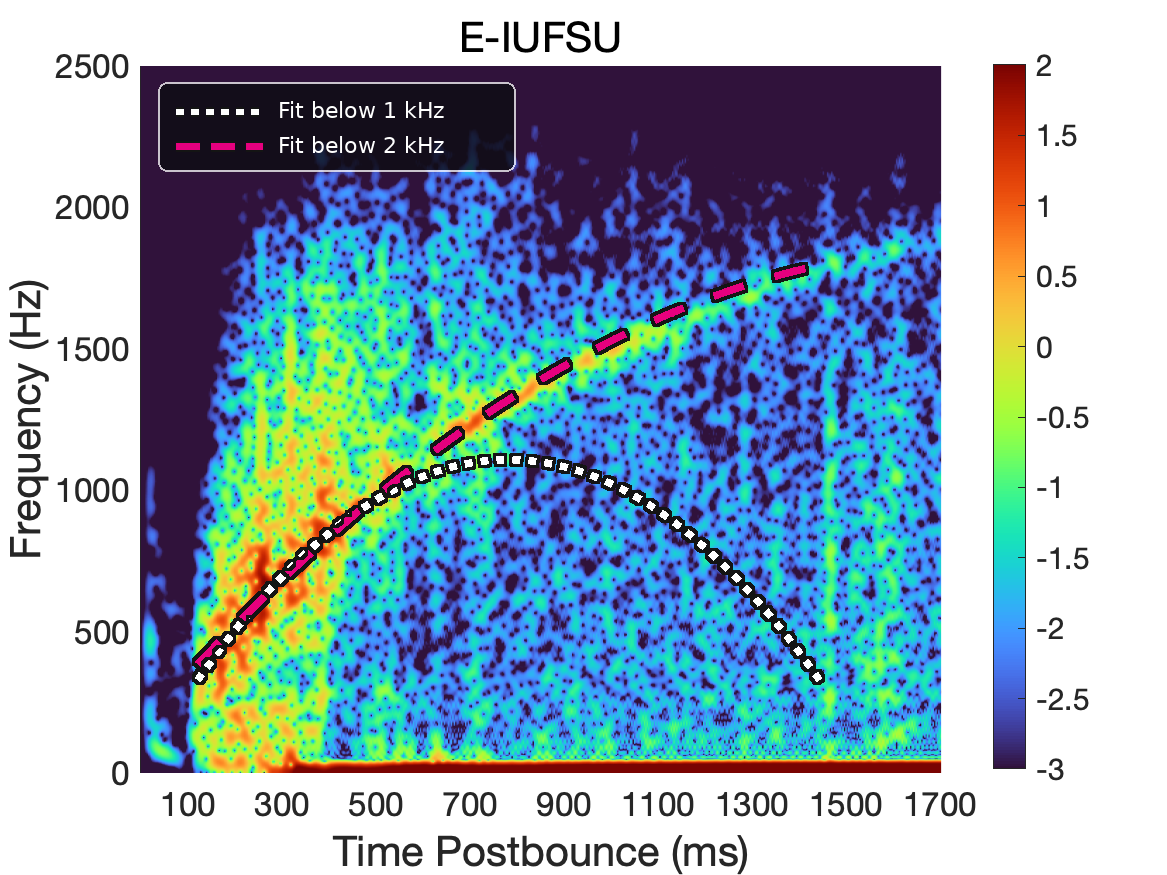}
		\includegraphics[width=0.30\textwidth]{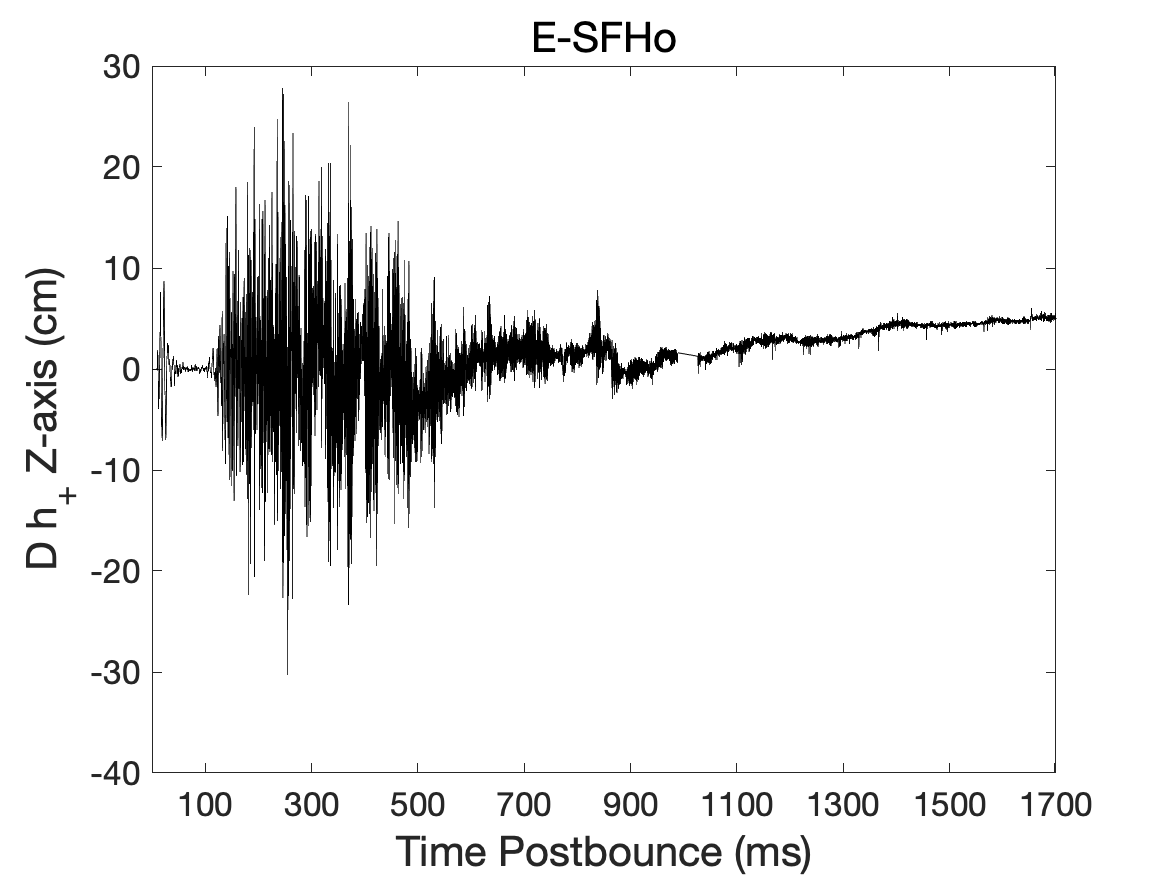}
		\includegraphics[width=0.35\textwidth]{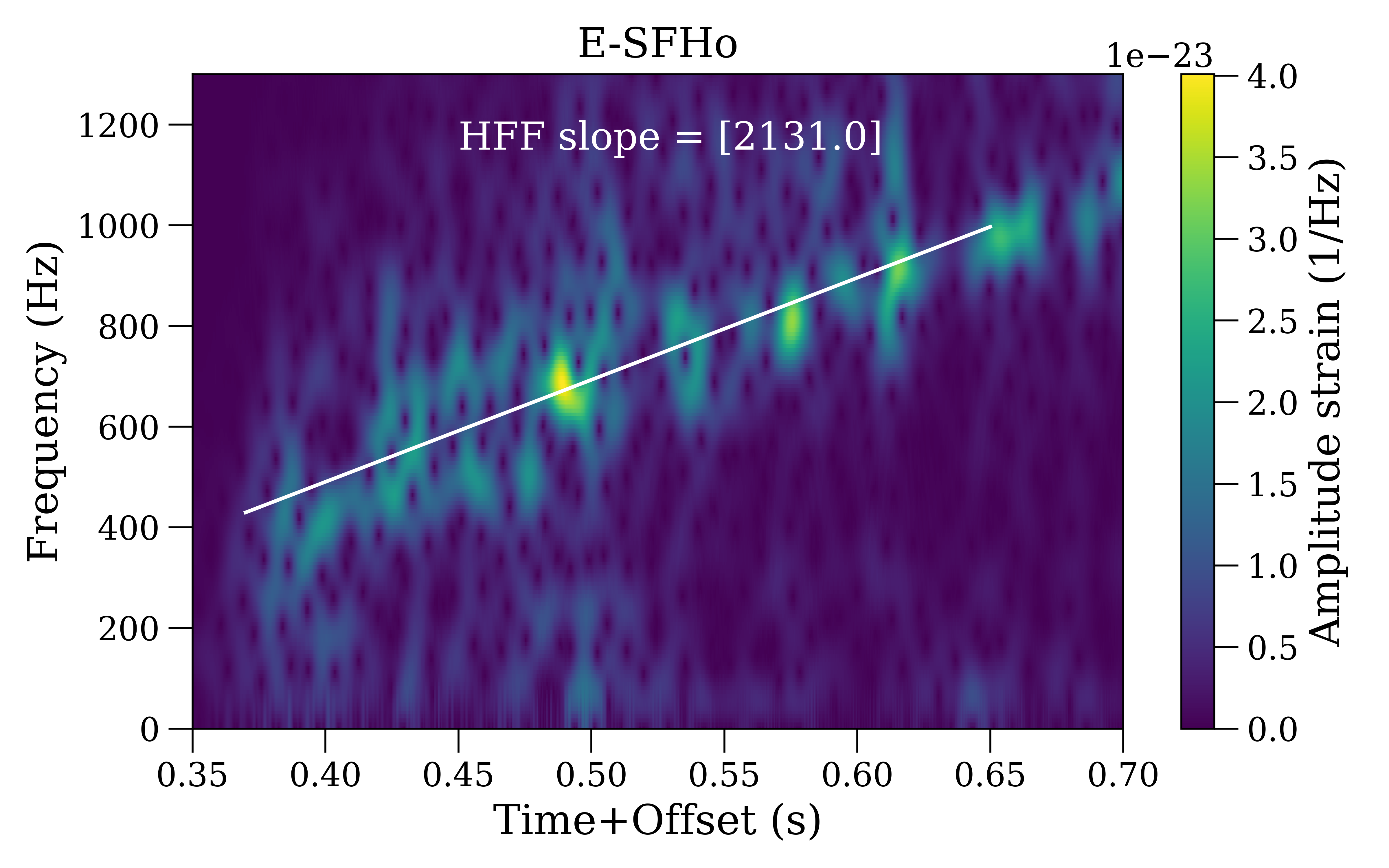}
		\includegraphics[width=0.30\textwidth]{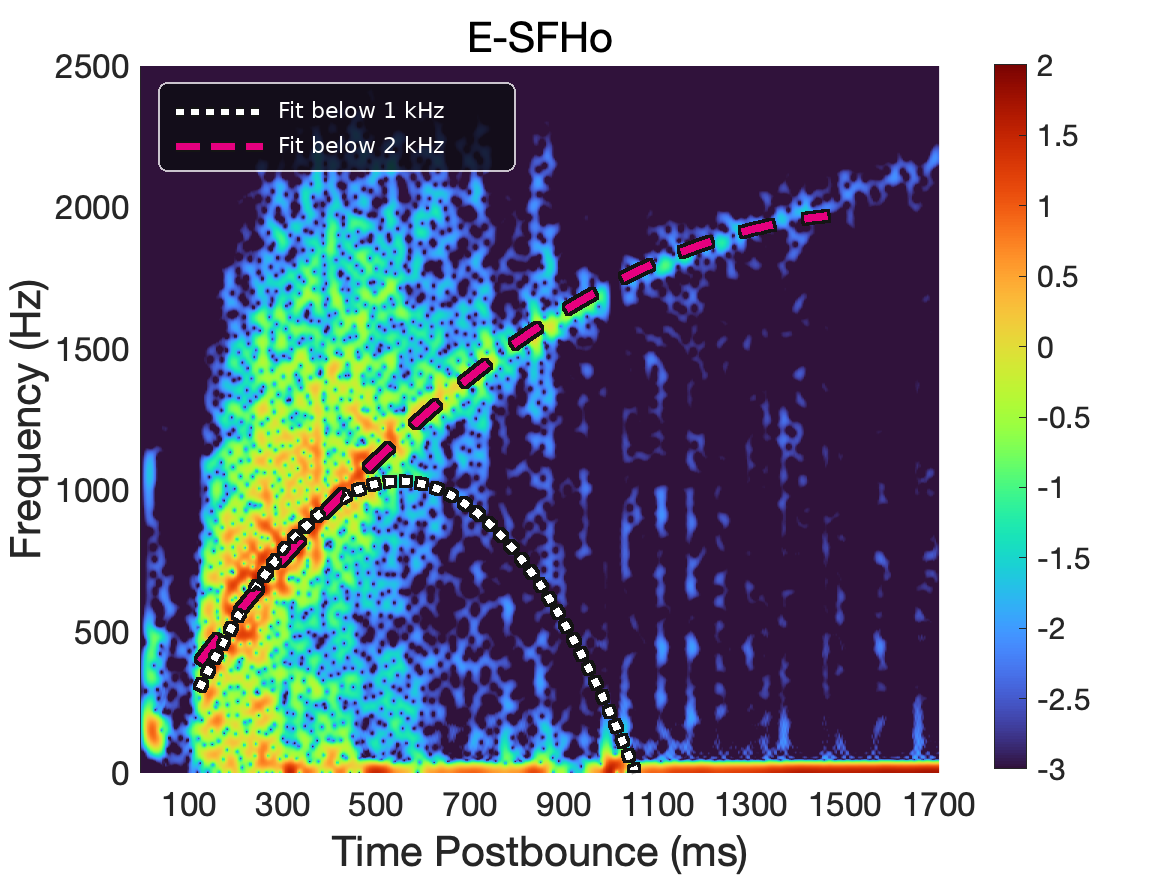}
		\includegraphics[width=0.30\textwidth]{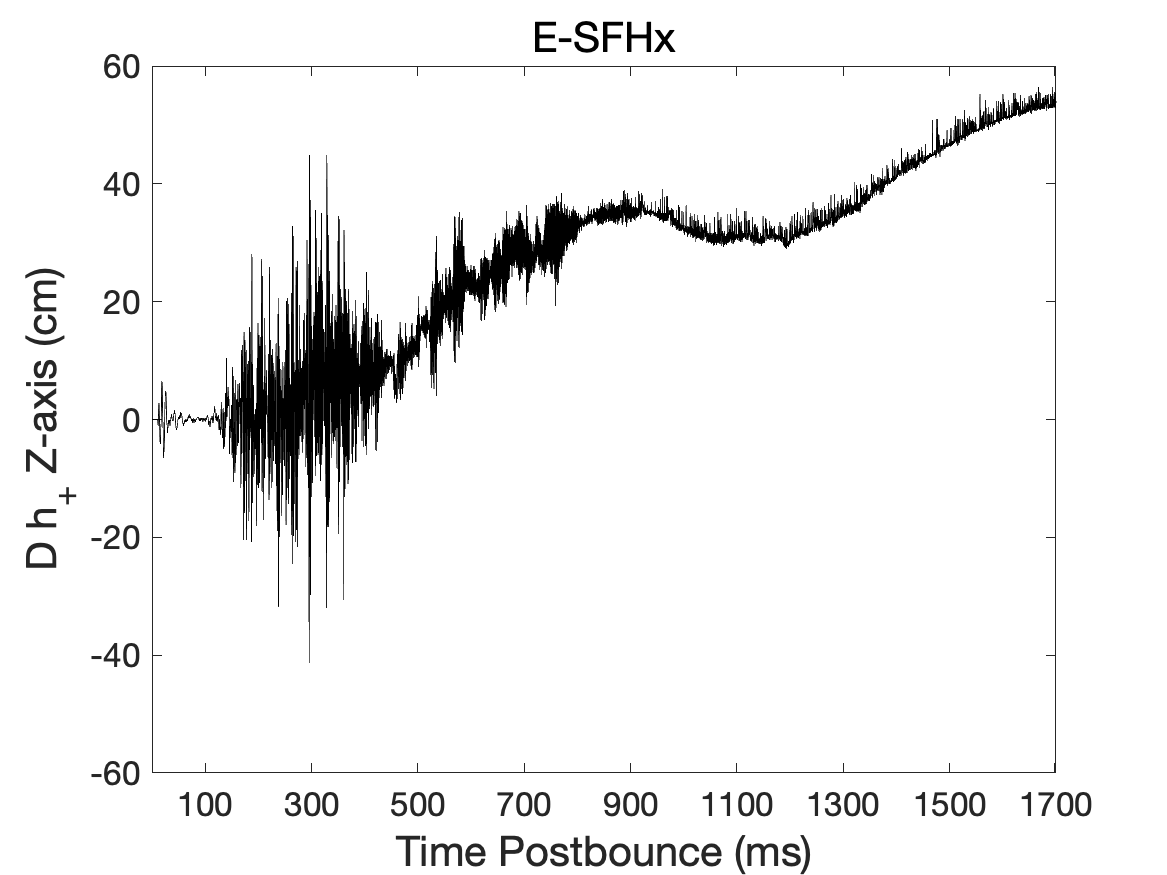}
		\includegraphics[width=0.35\textwidth]{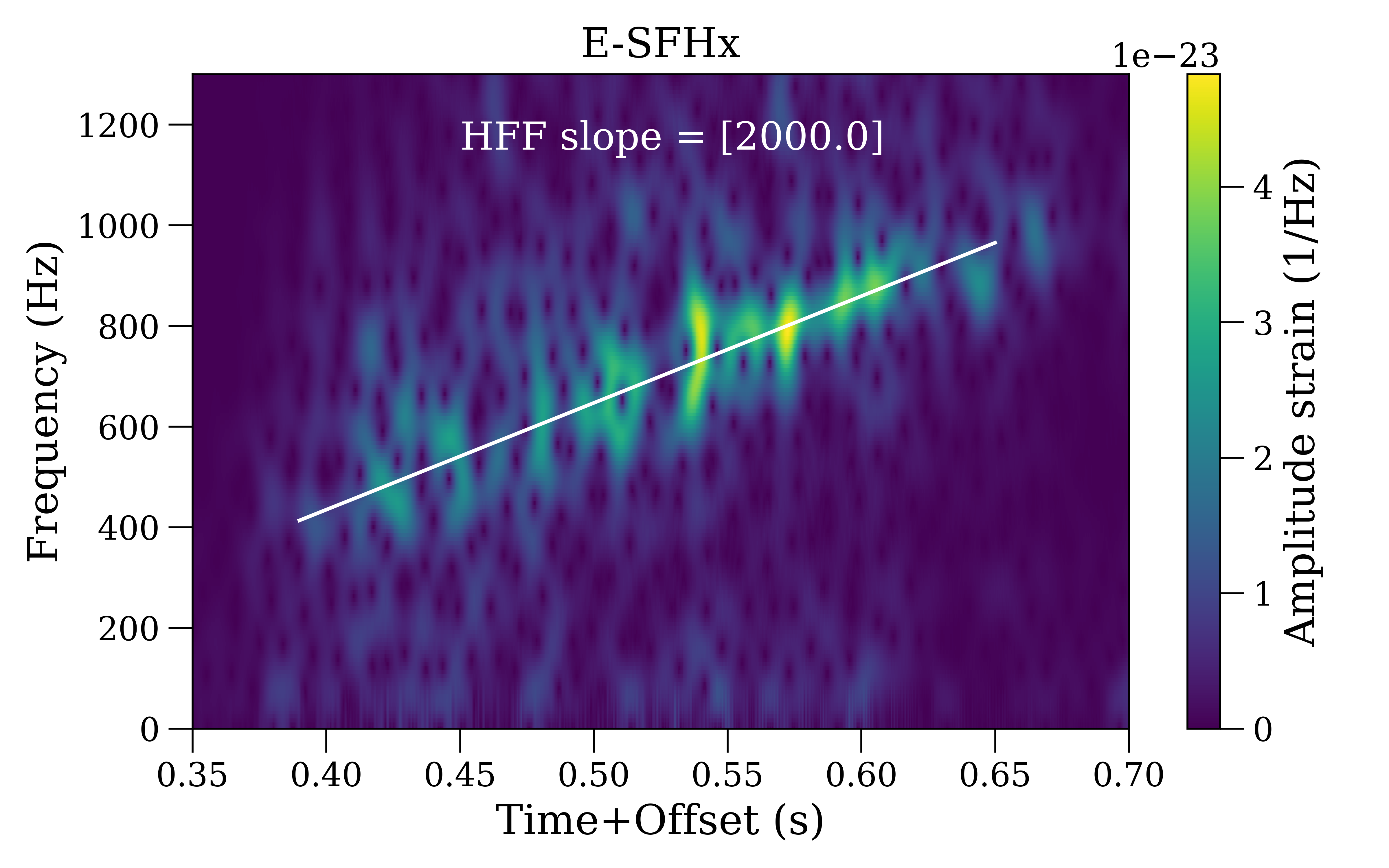}
		\includegraphics[width=0.30\textwidth]{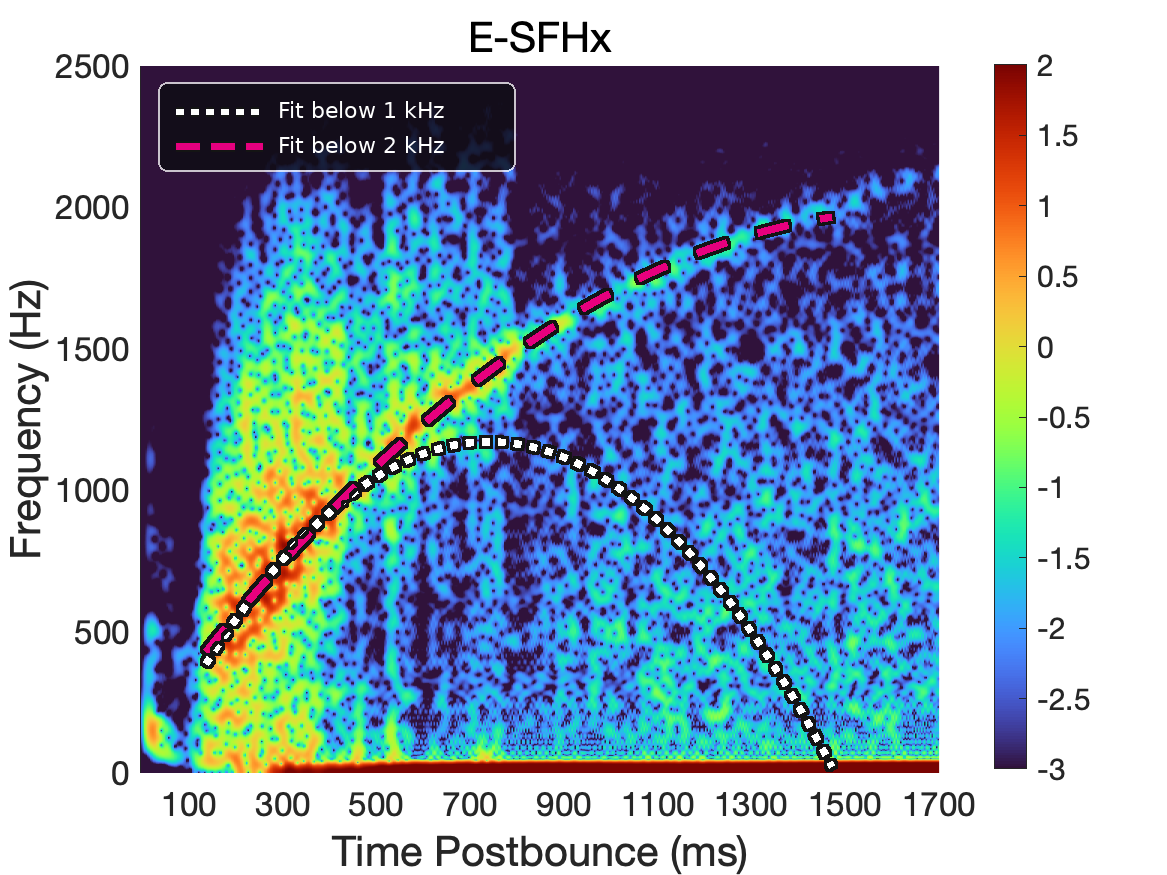}
		\caption{Strains and time--frequency representations of the CCSN GW signals from the axisymmetric \textit{CHIMERA} E-series simulations \cite{Bruenn_2020,Landfield2018} for the five EOS considered in \cite{Murphy_2024}: DD2, FSUgold, IUFSU, SFHo, and SFHx. The left column shows the simulated GW strain for each EOS. The middle column shows the HFF reconstructed after injection into real O3b interferometric noise, together with the estimated early-time linear slope following \cite{Casallas_1,Murphy_2024}. The horizontal coordinate in the middle column is a local reconstruction-time coordinate and is not referenced directly to core bounce. The right column shows the corresponding noiseless time--frequency representation. The white dotted and magenta dashed curves show second-order polynomial descriptions obtained using the HFF below $1~\mathrm{kHz}$ and $2~\mathrm{kHz}$, respectively. Because the fits use different data domains, they illustrate the dependence of the inferred evolution on the analyzed portion of the HFF rather than providing two measurements of an identical curvature coefficient.}
		\label{fig:signals}
	\end{figure}

	Figure~\ref{fig:signals} presents results from the axisymmetric two-dimensional \textit{CHIMERA} E-series simulations \cite{Bruenn_2020,Landfield2018}, which were designed to examine the influence of the EOS. Each simulation employs the same non-rotating, solar-metallicity, $15~M_{\odot}$ progenitor of \cite{Woosley_2007}, while varying the EOS among DD2, FSUgold, IUFSU, SFHo, and SFHx. Holding the progenitor, dimensionality, numerical framework, and neutrino treatment fixed provides a controlled setting in which differences in the HFF can be associated primarily with EOS-dependent changes in PNS structure and evolution.
	
	In Figure~\ref{fig:signals}, the left column shows the simulated GW strain for each EOS associated with the E-Series simulations. The middle column presents the HFF reconstructed after injection into real LVK O3b interferometric noise, together with the estimated early-time linear slope following \cite{Casallas_1,Murphy_2024}. The horizontal coordinate in these reconstruction panels is a local analysis-time coordinate and is not referenced directly to the physical core-bounce time. The right column shows the corresponding noiseless time--frequency representations and compares second-order polynomial descriptions obtained using two different frequency domains. The white dotted curves use the HFF below $1~\mathrm{kHz}$, whereas the magenta dashed curves use information extending below $2~\mathrm{kHz}$.
	
	Notice that the controlled nature of the E-series is its principal advantage for the comparison presented here. Nevertheless, these models are axisymmetric: two-dimensional dynamics modify turbulent flows and exclude non-axisymmetric motion, comparisons of 2D and 3D simulations show that the frequency evolution is already correct it is the amplitude that changes. The E-series should therefore be interpreted as demonstrating EOS sensitivity within a fixed numerical framework rather than establishing a universal quantitative mapping between HFF observables and EOS parameters. The three-dimensional simulations discussed later [see figure \ref{fig:model_comparison}], specifically \textit{Kuroda et al. 2016 \cite{Kuroda_2016}, model s15 (SFHx)} and, \textit{Radice et al. 2019 \cite{Radice_2019}, model s25} are instead used to illustrate the  morphological diversity expected when more realistic dynamics, progenitor structures, and numerical treatments are considered.
	
	\section{Unmodelled Gravitational-Wave Searches}
	\label{sec:cwb}
	
	In a GW search, the detector strain data are decomposed into localized wavelet components, or pixels. Pixels displaying statistically significant excess power \cite{Klimenko_2008} can then be combined into clusters whose coherence across the detector network is evaluated after accounting for relative arrival times and detector antenna responses. The resulting coherent cluster provides a reconstruction of the GW signal. Because this reconstruction is obtained from detector data containing noise, it does not necessarily recover every component of the original numerical waveform. This limitation is particularly important at high frequencies, where current interferometers are generally less sensitive and where the departure of the HFF from its early approximately linear evolution may become most apparent.
	
	Within cWB, this reconstruction problem is addressed through a coherent multi-detector analysis based on a multiresolution wavelet representation \cite{Klimenko_2005, Klimenko_2016, Klimenko_2022}. The standard cWB-2G algorithm typically employs the Wilson--Daubechies--Meyer (WDM) transform to represent the detector data at multiple time--frequency resolutions. This approach enables the analysis of transient signals with durations ranging from approximately $10~\mathrm{ms}$ to several seconds while balancing temporal and frequency resolution. The cWB analysis proceeds through two principal stages:
	
	\begin{enumerate}
		\item[(i)] \textit{Production stage}: Multi-detector strain data are transformed into WDM time--frequency representations at multiple resolutions. Pixels exhibiting statistically significant excess power are combined coherently across detectors using the relative arrival times and antenna-pattern responses evaluated over a predefined sky grid. Network quantities, including the maximum likelihood, coherent energy, residual or null energy, and correlation coefficients, are calculated to evaluate the consistency of the event across the detector network. Clusters satisfying the required coherence and energy-balance conditions are retained as candidate events.
		
		\item[(ii)] \textit{Post-production stage}: Candidate events are ranked, subjected to additional consistency requirements, and reconstructed. The reconstruction yields an estimate of the GW response in each detector, represented schematically by the signal vector $s[i]$, together with estimates of the source sky position, polarization, and coherent event statistics. These quantities are obtained within a constrained maximum-likelihood framework by comparing the reconstructed responses with the detector data $x[i]$.	
	\end{enumerate}
	
	\subsection{Signal-strength and coherent-reconstruction statistics}
	\label{sec:cwb_statistics}
	
	Two distinct signal-strength quantities are relevant to the discussion presented here: the optimal network signal-to-noise ratio (SNR) assigned to an injected waveform and the coherent event statistic reconstructed by cWB. These quantities should not be used interchangeably. For a waveform projected onto a network of $K$ detectors, the network SNR is defined as
	\begin{equation}
		\rho_{\mathrm{opt}}^{2}
		=
		4
		\sum_{k=1}^{K}
		\int_{f_{\min}}^{f_{\max}}
		\frac{
			\left|\widetilde{h}_{k}(f)\right|^{2}
		}{
			S_{n,k}(f)
		}
		df,
		\label{eq:optimal_network_snr}
	\end{equation}
	where $\widetilde{h}_{k}(f)$ is the Fourier transform of the injected waveform after projection onto detector $k$, and $S_{n,k}(f)$ is the corresponding detector-noise power spectral density. For a fixed waveform, source orientation, sky position, detector network, and noise spectrum, $\rho_{\mathrm{opt}}$ scales with the waveform amplitude and therefore approximately as $1/d$, where $d$ is the source distance.
		The cWB quantity denoted by $\rho_{\mathrm{cWB}}$ is instead reconstructed from the coherent event identified in the detector data. It is commonly referred to as the cWB effective correlated amplitude or coherent ranking statistic. In the standard cWB post-production notation, it is written as
	\begin{equation}
		\rho_{\mathrm{cWB}}
		=
		\sqrt{
			\frac{E_0}{K-1}
		},
		\label{eq:cwb_rho}
	\end{equation}
	where $K$ is the number of detectors in the network and $E_0$ denotes the corrected correlated-energy statistic evaluated over the time--frequency pixels belonging to the reconstructed event. Operationally, 
		the construction of $E_0$ depends on the coherent reconstruction and event-consistency conditions, $\rho_{\mathrm{cWB}}$ is not identical to the optimal matched-filter network SNR \cite{Klimenko_2005, Klimenko_2008}. The symbol $\rho$ used in the population-level analysis denotes $\rho_{\mathrm{cWB}}$. The terms `low $\rho$'' and `high $\rho$'' consequently refer to relative subsets of the reconstructed cWB effective correlated amplitude and not to fixed physical source-distance intervals. When an injected waveform is scaled before its addition to detector data, its assigned strength is instead denoted by $\rho_{\mathrm{opt}}$.
		
		\subsection{The cWB-XP reconstruction framework}
		
		The cWB-XP implementation considered in this work retains the coherent multi-detector framework of cWB-2G while incorporating the WaveScan reconstruction procedure \cite{Klimenko_2022}. WaveScan employs a bank of wavelet representations with different time--frequency resolutions and is intended to improve the representation of narrow and rapidly evolving transient structures in non-stationary detector noise.
		
		A multiresolution wavelet decomposition is constrained by the Heisenberg--Gabor trade-off between temporal and frequency resolution. Consequently, a representation that is well suited to one part of a transient signal may produce temporal or spectral leakage in another. The adaptive WaveScan procedure is designed to combine information from multiple resolutions so that the local representation can follow changes in signal morphology more effectively than a single fixed-resolution representation.
		
		In the context of CCSN signals, we investigate whether the reconstructed HFF can be followed beyond its early approximately linear portion and into the higher-frequency regime where changes in the rate of frequency evolution may become apparent. The reconstructed examples presented in the following section are used to illustrate the HFF morphology retained by cWB-XP
			(we leave though, to future work, the study of the optimization of this process).
	
	\section{The landscape of the High-Frequency Feature in CCSN Gravitational Waves for the next few years}
	The complete evolution of the HFF is currently being investigated through complementary approaches, including asteroseismology, numerical simulations, data analysis, and modal and spatial decompositions. Early-slope estimation represents only an initial step toward extracting the full information carried by the HFF as the community prepares for the first detection of GWs from a Galactic CCSN. A concrete example is provided by Murphy et al.\ \cite{Murphy_2024}, who investigated the sensitivity of the reconstructed early HFF slope to the EOS while keeping the progenitor and numerical framework fixed. Their study demonstrated that the early slope can retain EOS-dependent information after injection into LVK O3b interferometric data. This controlled comparison provides a foundation for evaluating the extent to which HFF observables may be recovered with current detectors and, in the future, with planned ground-based observatories such as the Einstein Telescope \cite{Punturo_2010} and Cosmic Explorer \cite{Cosmic_Explorer}.
	
	\begin{figure}[htb]
		\centering
		\includegraphics[width=0.35\textwidth]{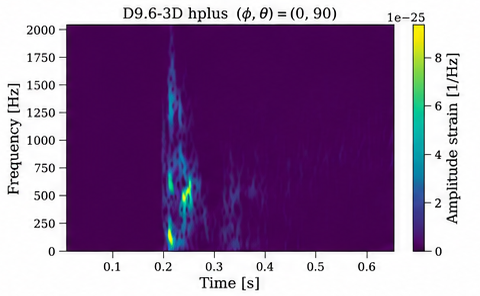}
		\includegraphics[width=0.30\textwidth]{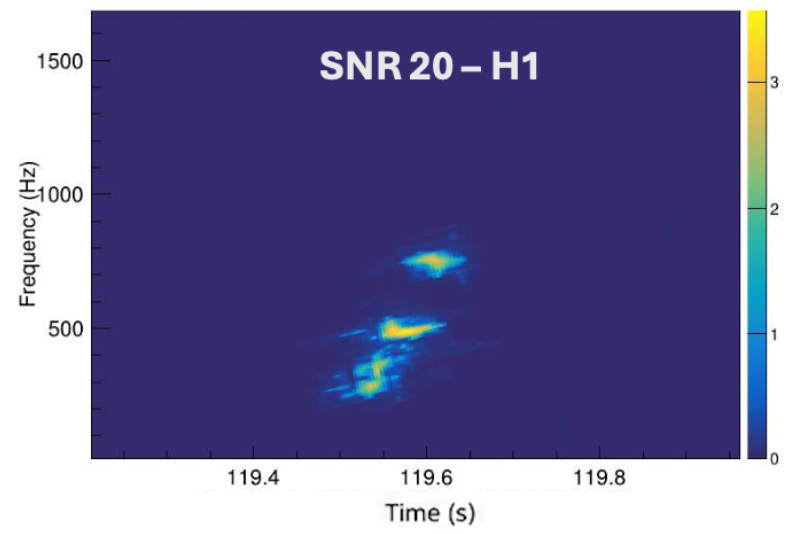}
		\includegraphics[width=0.30\textwidth]{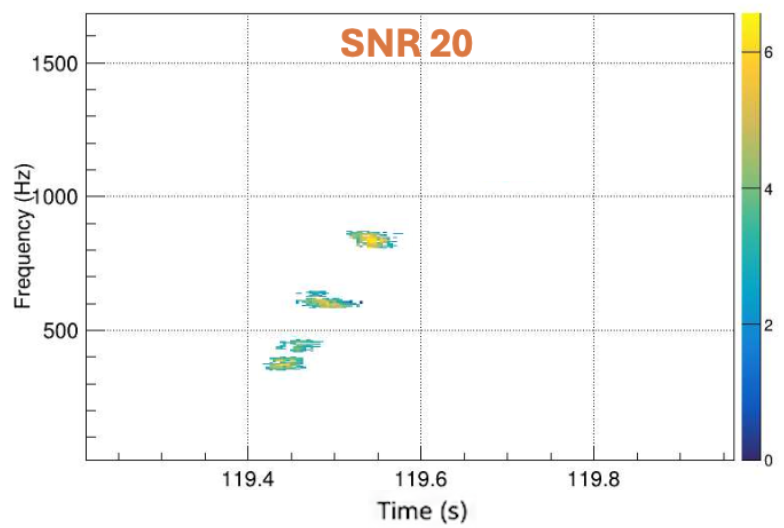}
		\includegraphics[width=0.35\textwidth]{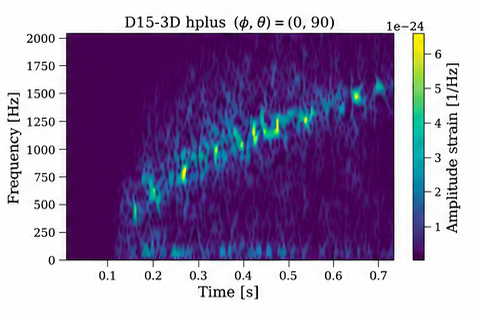}
		\includegraphics[width=0.30\textwidth]{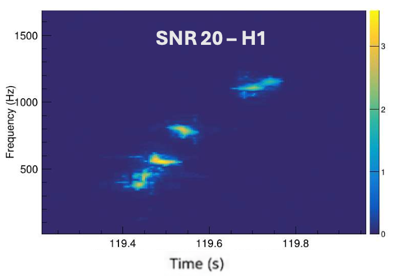}
		\includegraphics[width=0.30\textwidth]{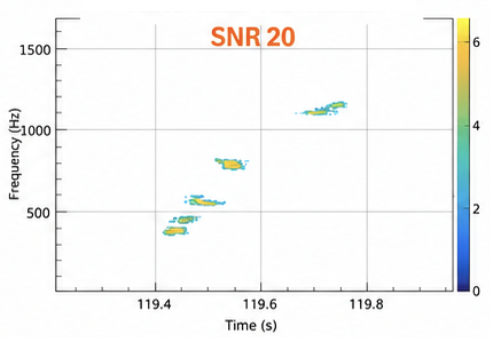}
		\includegraphics[width=0.35\textwidth]{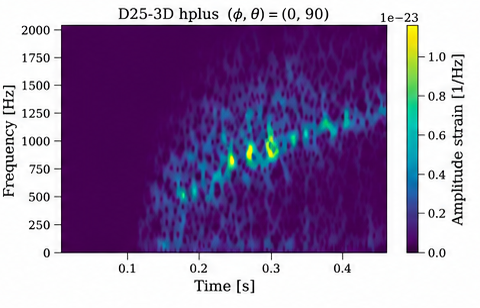}
		\includegraphics[width=0.30\textwidth]{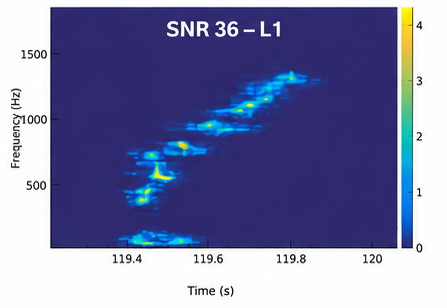}
		\includegraphics[width=0.30\textwidth]{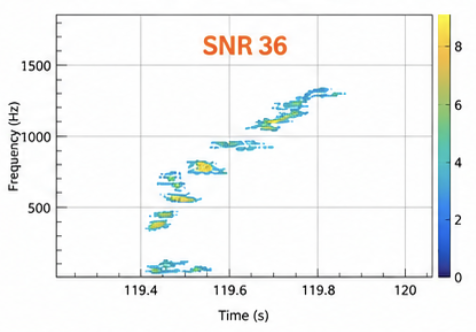}
		\caption{Numerical signals, detector injections, and representative cWB-XP time-frequency reconstructions for the three-dimensional D9.6-3, D15-3D, and D25-3D CCSN simulations of Mezzacappa et al. (2023)~\cite{Mezzacappa_2023}. The signals were injected into real O4c LVK interferometric noise. The reconstructed HFF appears as a rising time-frequency ridge. For sufficiently strong events—specifically D9.6 ($\mathrm{SNR} = 20$), D15 ($\mathrm{SNR} = 20$), and D25 ($\mathrm{SNR} = 36$)—the recovered ridge extends beyond its early, approximately linear portion and retains information about its subsequent nonlinear evolution, except for the selfinteracting D9.6 simulation, which exhibits a linear reconstruction at a fixed SNR of 20, the the models; D15 and D25 shows reconstructions beyond the early linear regime. The time coordinate shown in each reconstruction is referenced to the corresponding analysis window and does not represent time after core bounce.}
		\label{fig:mezza_3d}
	\end{figure}
	
\begin{figure}
		\centering
		\begin{subfigure}{\textwidth}
			\centering
			\includegraphics[width=0.39\textwidth]{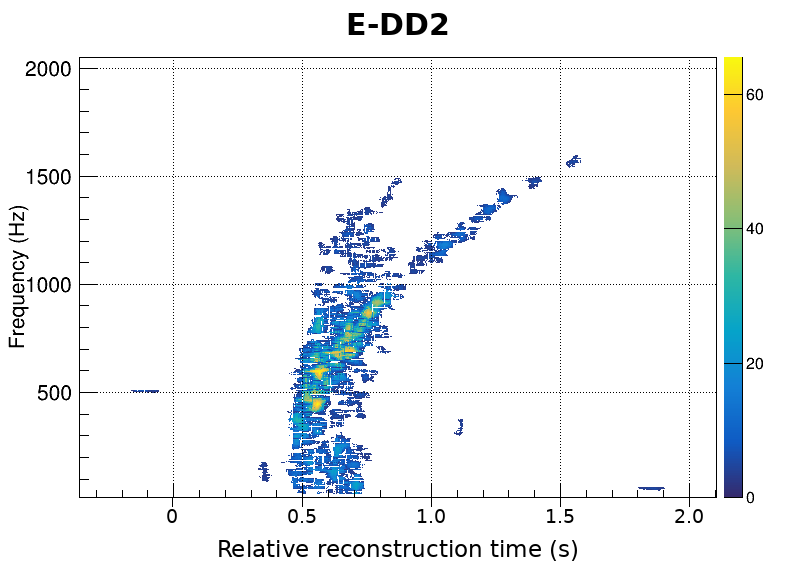}
			\includegraphics[width=0.39\textwidth]{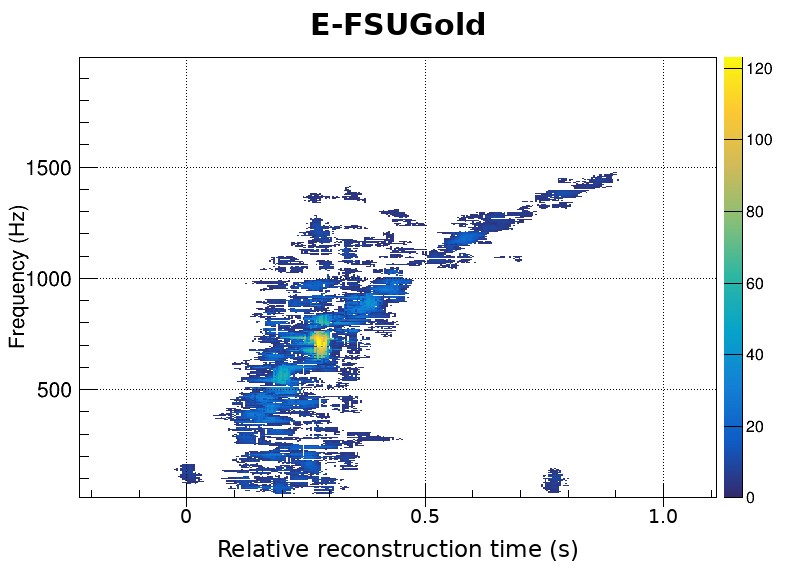}
			\includegraphics[width=0.39\textwidth]{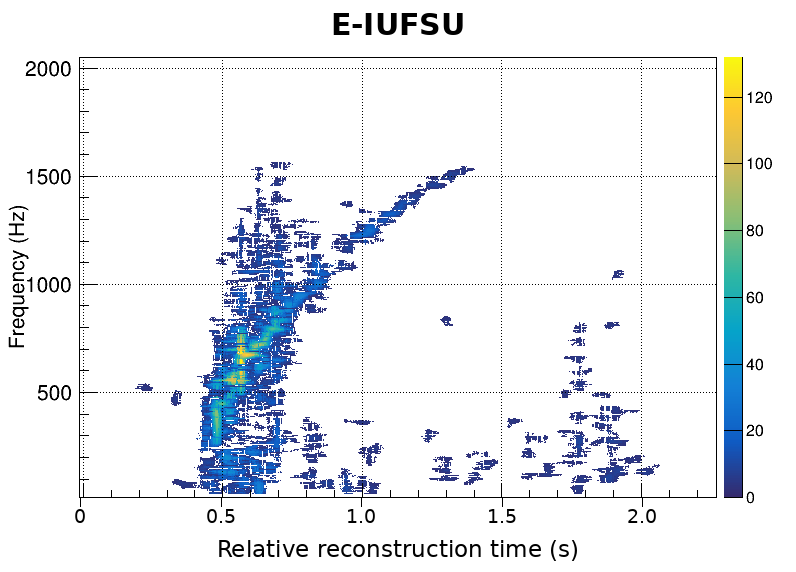}
			\includegraphics[width=0.39\textwidth]{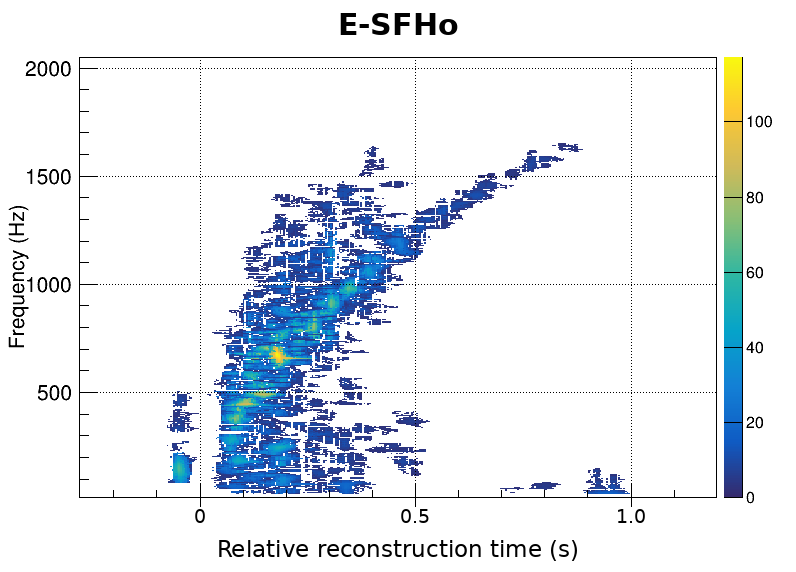}
			\includegraphics[width=0.39\textwidth]{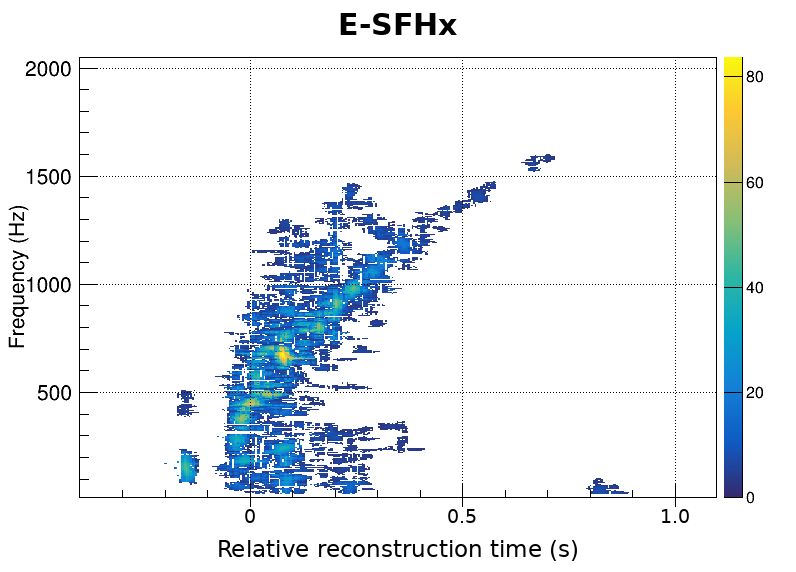}
			\caption{Reconstructions recovered by cWB-XP for the E-series simulations employing five different EOSs. The plots show a clear deviation from the linear behavior across all five considered EOSs. As previously reported in \cite{Murphy_2024}, a difference in the estimated slope was identified when signals were injected into O3b LVK noise following the methodology described in \cite{Casallas_1}.}
			\label{fig:e_series}
		\end{subfigure}
		
		\vspace{1em}
		
		\begin{subfigure}{\textwidth}
			\centering
			\includegraphics[width=0.41\textwidth]{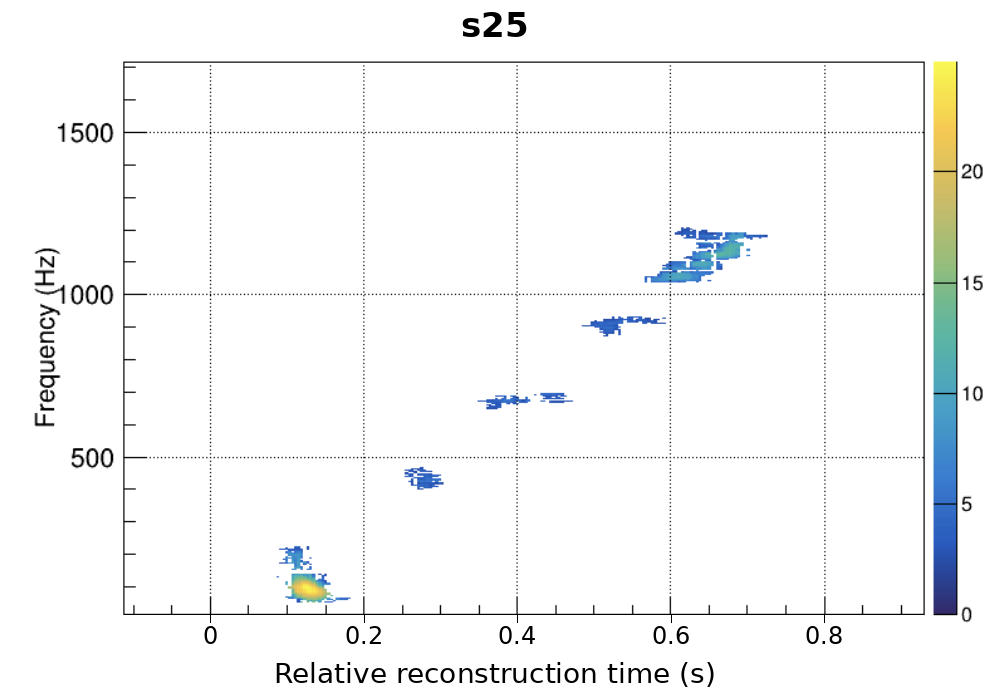}
			\includegraphics[width=0.41\textwidth]{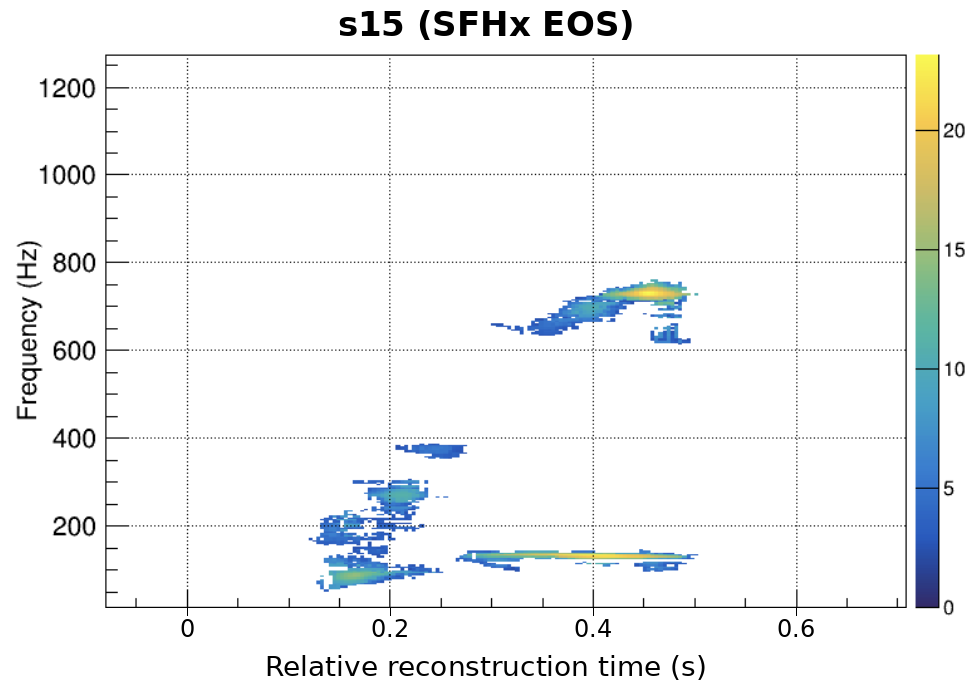}
			\caption{Comparison at $\mathrm{SNR} = 15$ of the reconstructed signals for the s25 model of Radice et al. (2019) and the s15 model of Kuroda et al. (2016), with the latter employing the SFHx EOS. The reconstructions recovered by cWB-XP illustrate how these two different 3D models exhibit distinct HFF evolutions: the s25 model remains approximately linear, whereas the s15 model displays a behavior that departs from the linear component.}
			\label{fig:model_comparison}
		\end{subfigure}
		
		Time-frequency reconstructions of CCSN GW signals obtained with cWB-XP after injection into real LVK interferometric noise at a controlled network $\mathrm{SNR} = 25$. Panel (a) presents the five \textit{CHIMERA} E-series simulations employing the DD2, FSUgold, IUFSU, SFHo, and SFHx EOSs. Panel (b) compares the s25 model of Radice et al. 2019 with the s15 model of Kuroda et al. 2016. In each panel, time is measured relative to a panel-specific reconstruction reference; thus, $t = 0$ serves as a plotting reference and does not represent the physical core-bounce time. The figure illustrates the diversity of reconstructed HFF evolution beyond its early, approximately linear portion, but does not by itself establish a statistically significant curvature or asymptotic frequency.
		\label{fig:recon_combined}
	\end{figure}

	A visual inspection of the CHIMERA E-series reconstructions reported by Murphy et al. (2024) suggests that cWB-XP is already capable of recovering HFF pixels beyond the early linear slope investigated in that work. Consistent with this, the CHIMERA E-series reconstructions in Figure~\ref{fig:recon_combined}(a) explicitly show that cWB-XP can retain portions of the HFF extending beyond its early, approximately linear segment. Figure~\ref{fig:recon_combined}(b) further illustrates that the later reconstructed morphology varies among CCSN GW simulations: while the model s25 of Radice et al. (2019) \cite{Radice_2019} remains comparatively close to a rising linear trajectory over the recovered interval, the model s15 of Kuroda et al. (2016) \cite{Kuroda_2016}, employing the SFHx EOS, shows a more pronounced change in its rate of frequency evolution when reconstructed at the same SNR. These visual differences are expected due to the different physical treatments implemented in the numerical simulations, including progenitor mass, metallicity, rotation, microphysics, gravity, and overall model sophistication among others. As will be demonstrated below using the 3D waveforms of Mezzacappa et al. 2023 \cite{Mezzacappa_2023}, cWB-XP effectively recovers these advanced pixels associated with the true HFF curvature.
	
	The next methodological step is to investigate whether cWB-XP can reconstruct the HFF beyond its early, approximately linear evolution. For this purpose, we consider the three-dimensional D9.6-3, D15-3D, and D25-3D CCSN waveforms presented by Mezzacappa et al. \cite{Mezzacappa_2023}. These signals were injected into real O4c LVK interferometric noise and subsequently reconstructed using cWB-XP. We examine whether the reconstructed time-frequency tracks retain information from the portion of the HFF extending beyond the early, approximately linear regime.
		Figure~\ref{fig:mezza_3d} compares (i) in the left panel, the numerical signals; (ii) in the middle panel, their detector injections in O4c LVK data; and (iii) in the right panel, the corresponding cWB-XP reconstructions. The reconstructed events serve to visually demonstrate that cWB-XP can recover the early rising portion of the HFF and, for sufficiently strong events, retain part of its subsequent nonlinear evolution. Recovering this later portion is more challenging because it generally occurs at higher frequencies, where the sensitivity of current interferometers is reduced.
	
	To examine how this reconstruction depends on signal strength, we divide the recovered events according to the cWB network statistic ($\rho_{\mathrm{cWB}}$), which provides a dimensionless measure of the coherent strength of a reconstructed event. For each simulation group $g\in\{\mathrm{D15\mbox{-}3D},\mathrm{D25\mbox{-}3D},
		\mathrm{D9.6\mbox{-}3}\}$
		the low- and high-($\rho_{\mathrm{cWB}}$) populations are defined as the lower and upper quartiles of the corresponding group-specific distribution:
	\begin{equation}
		L_g=
		\left\{
		i:\rho_{ig}\leq q_{25,g}
		\right\},
		\qquad
		H_g=
		\left\{
		i:\rho_{ig}\geq q_{75,g}
		\right\},
		\label{eq}
	\end{equation}
	where ($q_{25,g}$) and ($q_{75,g}$) denote the 25th and 75th percentiles of ($\rho_{\mathrm{cWB}}$) for simulation group (g). The intermediate 50\% of the reconstructed events is excluded from this comparison. The group-specific thresholds are reported in Table~\ref{tab:SNR}.
	
	\begin{table}
		\centering
		\caption{Group-specific values of the cWB network statistic used to define the low and high $\rho_{\mathrm{cWB}}$ populations. Events below $q_{25,g}$ form the low $\rho_{\mathrm{cWB}}$ population, whereas events above $q_{75,g}$ form the high-$\rho_{\mathrm{cWB}}$ population.}
		\label{tab:rho_quartiles}
		\begin{tabular}{lcc}
			\toprule
			\textbf{Simulation} &
			$\boldsymbol{q_{25,g}}$ &
			$\boldsymbol{q_{75,g}}$ \\
			\midrule
			D15-3D & 12.80 & 25.95 \\
			D25-3D & 11.14 & 40.90 \\
			D9.6-3 & 11.08 & 30.18 \\
			\bottomrule
			\label{tab:SNR}
		\end{tabular}
	\end{table}
	
\begin{figure}[htb]
		\centering
		\includegraphics[width=1.0\textwidth]{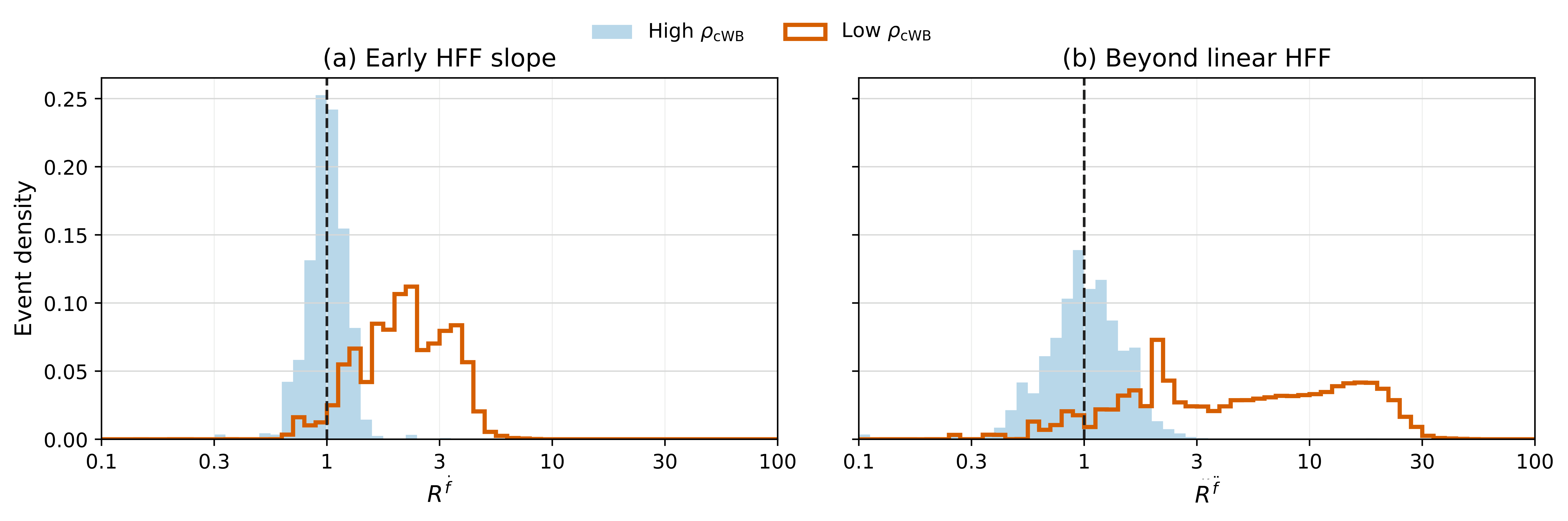}
		\caption{Event comparison of the first- and second-order HFF evolution for the Mezzacappa et al. 2023 D9.6-3, D15-3D, and D25-3D simulations. The blue filled and orange outlined histograms represent the high- and low-$\rho_{\mathrm{cWB}}$ populations, respectively, as defined by the group-specific upper and lower quartiles. Each reconstructed event contributes one value to each panel. The quantities are normalized by the corresponding high-$\rho_{\mathrm{cWB}}$ median within each simulation, after which the three simulation groups are combined with equal total weight. Panel (a) shows $R^{\dot f}$, the normalized first derivative that determines the early approximately linear HFF evolution in equation~\eqref{eq:hff_frequency_derivative}. Panel (b) shows $R^{\ddot f}$, the normalized second derivative equation~\eqref{eq:hff_second_frequency_derivative} that determines the quadratic departure from that linear evolution. The horizontal axes are logarithmic, and the dashed line at unity marks the corresponding high-$\rho_{\mathrm{cWB}}$ reference median.}
		\label{fig:population}
	\end{figure}

	Each selected reconstruction contributes a single value for the early HFF slope and another characterizing the evolution beyond the linear regime. These two quantities correspond directly to the first and second order terms in the local expansion introduced in equation~\eqref{eq:hff_local_expansion}: $\dot f(t_s)$ determines the locally linear evolution of the HFF, whereas $\ddot f(t_s)$ determines how the slope changes with time and therefore describes the departure from linearity. The histograms in Figure~\ref{fig:population} consequently represent distributions of these event-level quantities obtained from cWB reconstructions, rather than distributions of individual time--frequency pixels. To compare the three simulations (D9.6-3, D15-3D, and D25-3D) on a common scale, each quantity is normalized by the median value obtained from the high-$\rho_{\mathrm{cWB}}$ population, defined using equation~\ref{eq:cwb_rho}, of the corresponding simulation. For a per-event quantity $X_{ig}$, we define
	\begin{equation}
		R^{X}_{ig}\equiv\frac{|X_{ig}|}{X^{\mathrm{ref}}_{g}},
		\qquad
		X^{\mathrm{ref}}_{g}\equiv
		\underset{j\in H_g}{\operatorname{med}}|X_{jg}|,
		\label{eq:normalized_hff_quantity}
	\end{equation}
	where $i$ labels the reconstructed event, $g$ labels the simulation, and $H_g$ denotes its high-$\rho_{\mathrm{cWB}}$ population. We use $R^{\dot f}$ for the normalized first derivative in Figure~\ref{fig:population}(a) and $R^{\ddot f}$ for the normalized second derivative in Figure~\ref{fig:population}(b); the event and simulation indices are omitted from the axis labels for readability. In the notation used for the quadratic term of equation~\eqref{eq:hff_second_frequency_derivative}, $\kappa=\tfrac{1}{2}\ddot f(t_s)$. Because the constant factor $1/2$ appears in both the event value and its reference median, it cancels in the normalized ratio, so that $R^{\kappa}=R^{\ddot f}$. Thus, the change from $R^{\kappa}$ to $R^{\ddot f}$ is only a change of notation and does not modify the plotted distributions. After normalization, the three simulation groups are combined with equal total statistical weight, preventing the larger D15-3D and D25-3D reconstruction samples from dominating the D9.6-3 sample.
	
	For each Mezzacappa et al. 2023 simulation, the first and second derivative values reconstructed for an individual event are divided by the corresponding medians obtained from the high-$\rho_{\mathrm{cWB}}$ events of that same simulation. Consequently, $R^{\dot f}=1$ means that the reconstructed first derivative equals the high-$\rho_{\mathrm{cWB}}$ reference slope for that simulation, while $R^{\ddot f}=1$ means that the reconstructed second derivative equals its corresponding high-$\rho_{\mathrm{cWB}}$ reference value. The dashed vertical line at unity therefore does not represent a single absolute value of $\dot f$ or $\ddot f$ shared by the three simulations. The absolute references are calculated separately for D9.6-3, D15-3D, and D25-3D before the normalized distributions are combined. In panel (a), the high-$\rho_{\mathrm{cWB}}$ distribution is narrow and concentrated near $R^{\dot f}=1$, showing that the coefficient of the linear term in equation~\eqref{eq:hff_local_expansion} is recovered comparatively consistently for events with stronger coherent reconstructions. The low-$\rho_{\mathrm{cWB}}$ distribution is broader and shifted toward values above unity, indicating that the recovered first derivative becomes more variable and systematically displaced from its reference as the coherent signal strength decreases. In panel (b), the low-$\rho_{\mathrm{cWB}}$ distribution of $R^{\ddot f}$ extends over a substantially wider range. The coefficient governing the quadratic term in equation~\eqref{eq:hff_local_expansion} is therefore more sensitive to signal strength and detector noise than the coefficient of the linear term. This hierarchy is physically expected because the second derivative depends on recovering the later and generally higher-frequency portion of the HFF. Figure~\ref{fig:population} thus provides a concrete observational interpretation of equation~\eqref{eq:hff_local_expansion}: cWB-XP recovers the first-order evolution more robustly, while information about the second-order evolution is retained primarily in sufficiently strong reconstructions.
	
	\section{Conclusions and Outlook}
	The  HFF  remains a promising observable connecting CCSN GW signals with the evolving structure of the PNS.
		. Extending the analysis beyond the approximately linear regime may provide complementary information about changes in the PNS regarding contraction, accretion, and oscillation-mode structure. Building on visual indications from 2D CHIMERA reconstructions, the 3D waveforms presented here motivate this extension by demonstrating that information from the later HFF can remain accessible in sufficiently strong events, while curvature-related quantities are appreciably more sensitive to reconstruction variability at lower network strength.
	
	Several methodological challenges must nevertheless be resolved before HFF curvature can be treated as a quantitative inference observable. A reconstructed bend may describe genuine nonlinear evolution of a single mode, a transition between different mode branches, finite-interval fitting effects, detector noise, or properties of the adaptive time--frequency reconstruction. A validated analysis will therefore require a robust estimator to determine how physical parameters of the source—such as the EOS, progenitor mass, rotation rate (and associated effects like centrifugal support and Coriolis forces), magnetic fields, metallicity, and simulation dimensionality—affect both the early and late stages of HFF evolution. Furthermore, a corresponding astrophysical interpretation of such an analysis is required to unlock deeper insights into PNS physics revealed through its GW emission.
	
	The present work therefore identifies HFF curvature not as an already established measurement, but as a well-motivated direction for future CCSN GW research. Combining improved time--frequency reconstruction, controlled numerical simulations, and advanced mode identification may ultimately extend the information obtained from a future Galactic CCSN beyond the early HFF slope and toward a more complete, high-fidelity description of the evolving PNS.
	
	\FloatBarrier 
	
	\section*{Acknowledgements}	
	A.C. and M.S. acknowledge Polish National Science Centre Grant No. UMO-2024/03/1/ST9/00005, and the Polish National Agency for Academic Exchange within Polish Returns Programme Grant No. BPN/PPO/2023/1/00019. M.S. acknowledges also Polish National Science Centre Grant No. UMO-2023/49/B/ST9/02777. M.Z. is supported by the National Science Foundation Gravitational Physics Experimental and Data Analysis Program through awards PHY-2110555 and PHY-2405227. A.M. acknowledges support from National Science Foundation Gravitational Physics Theory under awards PHY-1806692, 2110177, 2409148.

	\bibliographystyle{unsrt}
	\bibliography{references}
	
\end{document}